\pdfoutput=1
\documentclass[fleqn,10pt]{./lib/SelfArx} 

\usepackage[english]{babel} 
\usepackage{tabularx}
\usepackage{multirow}
\usepackage{lipsum} 

\usepackage[utf8]{inputenc}
\usepackage{textcomp}
\usepackage{gensymb}
\usepackage{subfig}
\usepackage{url}

\definecolor{color1}{RGB}{0,0,90} 
\definecolor{color2}{RGB}{0,20,20} 

\usepackage{hyperref} 
\hypersetup{hidelinks,colorlinks,breaklinks=true,urlcolor=color2,citecolor=color1,linkcolor=color1,bookmarksopen=false,pdftitle={Title},pdfauthor={Author}}

\JournalInfo{arXiv e-print - Computer Science - Human-Computer Interaction} 
\Archive{Original research article} 

\PaperTitle{Sunny Pointer: Designing a mouse pointer for people with peripheral vision loss} 

\Authors{Maxime Ambard\textsuperscript{1}*} 
\affiliation{\textsuperscript{1}\textit{LEAD CNRS UMR 5022, Université Bourgogne Franche-Comté}} 
\affiliation{*\textbf{Corresponding author}: maxime.ambard@u-bourgogne.fr} 

\newcommand{\keywordname}{Keywords} 

\Abstract{
We present a new mouse cursor designed to facilitate the use of the mouse by people with peripheral vision loss. The pointer consists of a collection of converging straight lines covering the whole screen and following the position of the mouse cursor. We measured its positive effects with a group of participants with peripheral vision loss of different kinds and we found that it can reduce by a factor of 7 the time required to complete a targeting task using the mouse. Using eye tracking, we show that this system makes it possible to initiate the movement towards the target without having to precisely locate the mouse pointer. Using Fitts' Law, we compare these performances with those of full visual field users in order to understand the relation between the accuracy of the estimated mouse cursor position and the index of performance obtained with our tool.
}

\begin{document}

\flushbottom 

\maketitle 

\tableofcontents 

\thispagestyle{empty} 


\section*{Introduction} 

\addcontentsline{toc}{section}{Introduction} 

Peripheral vision loss (PVL), also known as "tunnel vision", is a disability in which a person’s visual field (VF) is restricted to a small centered portion of its normal size \cite{Rosenholtz_2016}. It may be caused by various diseases such as retinitis pigmentosa or glaucoma, which was identified in 2002 as the second cause of blindness worldwide according to the World Health Organization \cite{Resnikoff_2004} with a projection of 100 millions of affected people in 2040 \cite{Tham2014}. People with PVL encounter several major difficulties that can severely affect their personal and professional lives \cite{Quaranta_2016, Ramulu_2009, evans2009}. Despite its low resolution \cite{westheimer1982}, peripheral vision provides important information about the environment \cite{thorpe2001, larson2009} and can guide the gaze for high-resolution inspection with the fovea during visual search \cite{geisler2006, hooge1999, coeckelbergh2002} and can provide online information in the control of movement direction \cite{khan2004}. The use of a computer is problematic for people with PVL \cite{Jacko_1998, Jacko_2000_HCI}.

One of the main difficulties for these people is the use of the mouse, which remains one of the most commonly used tools for the selection of interactive items spatially distributed on a computer screen. The problem is that the surface of a screen of 50.9cm in width and 28.6cm in height placed 70cm away from the user is approximately 110 times bigger than what a person with a centered visual field (VF) reduced to 1.5$^\circ$ radius sees. This decreases the speed of the PVL computer user and require important cognitive resources that can bring about mental fatigue that can become unbearable. A system specifically designed to improve the accessibility of the computer mouse for people with PVL is thus of great import in improving their quality of life.

Due to their restricted field of view, the difficulties encountered by users with a PVL impairment to click on a target can be separated in three successive steps. First, the user has to localize the area where he or she wants to click on. This depends on many parameters such as its color, its position, its size and the user prior knowledge of the graphical interface. Second, the user has to retreive where the mouse pointer is located prior to moving it toward a target on the screen. And third, PVL users have to control the mouse pointer trajectory towards the target which is difficult since they can not simultaneously see the target and use their peripheral vision to control the mouse pointer trajectory by updating their movements \cite{proteau2000}. In this paper we describe a new assistive solution to help poeple with PVL to use their mouse during the two last steps (i.e. the initial localization of the mouse pointer and its trajectory control). 

As described in \cite{Fraser_2000}, such an assistive technology may act on four dimensions. The first one is the perceptual channel used to assist the user. It may be visual, auditory or tactile. The second dimension describes whether knowledge of its context is required by the assistive system. In the case of a mouse cursor, this corresponds to whether or not the mouse cursor has preliminary knowledge of the targets that can be accessed on the screen. The third refers to the phase of operation that is eased by the system. In our case, this may be the localization of the pointer or the target, the move of the pointer toward the target or the click on the precise position of the target. The fourth dimension, pervasiveness, describes whether the system is turned on occasionally or permanently. 

Referring to the two first criteria mentioned, the pointer design that we present belongs to the category of visual and context-agnostic technologies. Previously developed assistive tools to ease the use of the mouse by the visually impaired are already included in this category \cite{liu2018}. The most obvious members of this category are pointer magnification tools which are now integrated into every common operating system. They consist of tools which increase the size of the pointer, draw large circles around the pointer or draw a visual trail materializing the recent pointer moves \cite{Fraser_2000, Baudisch2003}. Although use of these tools can undoubtedly help to localize the mouse pointer, people with severe PVL still have difficulty since these systems do not help to simultaneously see the pointer and the target and thus provide no assistance during the move of the pointer toward the target. Moreover, the proper setting value for the size of these visual cues results from a trade-off between the visibility of the pointer and the visibility of the rest of the interface. These tools are thus designed more for people with low visual acuity than for people with PVL \cite{Chiang_2005, Jacko_2000}.

The most common tool used by PVL users seems to be currently \textit{ZoomText} \cite{ZoomText} that draws a big crosshair composed of one vertical line and one horizontal line that intersect at the mouse position. Despite its qualities, seeing an horizontal line (resp. vertical) superimposed to the target does not allows the user to directly know if the pointer is at the right or at the left (resp. above or below) the target. The user has to move the pointer, potentially in the opposite direction of the target, to disambiguate the information.

\textit{Color Eyes} is another system that draws a stylized pair of eyes on the screen which continually gaze toward the mouse pointer and encode its distance by means of a color code \cite{Kline_1995}. Using this tool, the user can have an approximate idea of the localization of the mouse pointer and thus search for it in a reduced portion of the screen. However, in order to localize the pointer, this assistive technology requires the user to first look at an additional visual component, a process which can perturb the memorization of the localization of the target.

A new version of the tool called \textit{RPMouse} has been recently released. This tool draws a line from the top left corner of the screen to the mouse pointer position. Compares to standard a regular use of the mouse pointer, the user can more easily localize the mouse pointer without the need to visually scan the entire screen. To move the pointer towards the target, the user has first to look at the top left corner of the screen, to visually follow the line to its end in order to localize the cursor, then to move the cursor in an estimated direction of the target until both the cursor and the target enter the field of view. 

Another solution currently used by people with PVL consists in placing the pointer to a predefined position on the screen, for example by pressing a combination of hotkeys such as provided by the software \textit{AutoHotkey} \cite{AutoHotKey} or pressing a specific button on the mouse \cite{Hollinworth2011}. Other people manually places the pointer on the top left of the screen after each click. With these technics the mouse pointer is thus avoided, but they might place the pointer in a suboptimal initial position and it does not help in the case where the user becomes confused about the position of the pointer during its move toward a target. 

Despite all these tools, the use of the mouse pointer is still very difficult for the people with PVL. For these special-needs users, we have thus developed a mouse pointer called \textit{Sunny Pointer} that speeds up the use of the mouse to click on a target without ever losing sight of it as people who are able to use their full visual field would do. It can be freely downloaded at the following website \cite{sunny_pointer_url}. In this work, we first describe our pointer design and experimental methods. We then present a first experiment with six PVL users showing that our tool can decrease the time to complete the task by a factor up to $7$. In a second experiment, we used eye tracking with an expert peripheral vision loss participant that reveals that the pointer can be moved within the peripheral visual field towards the target while keeping the gaze on the target. In order to identify the characteristics of our pointer that could be tuned so as to obtain performance in participants with PVL close to those of a person with normal vision using a standard pointer, we compare, in a third experiment, the performance obtained by people with normal vision but with simulated PVL, with and without the Sunny Pointer.

\section{Materials and Methods}

\subsection{General Description of the Experiments}
\label{exp_descr}

\begin{figure*}[ht]\centering 
\subfloat{{\includegraphics[width=0.47\linewidth]{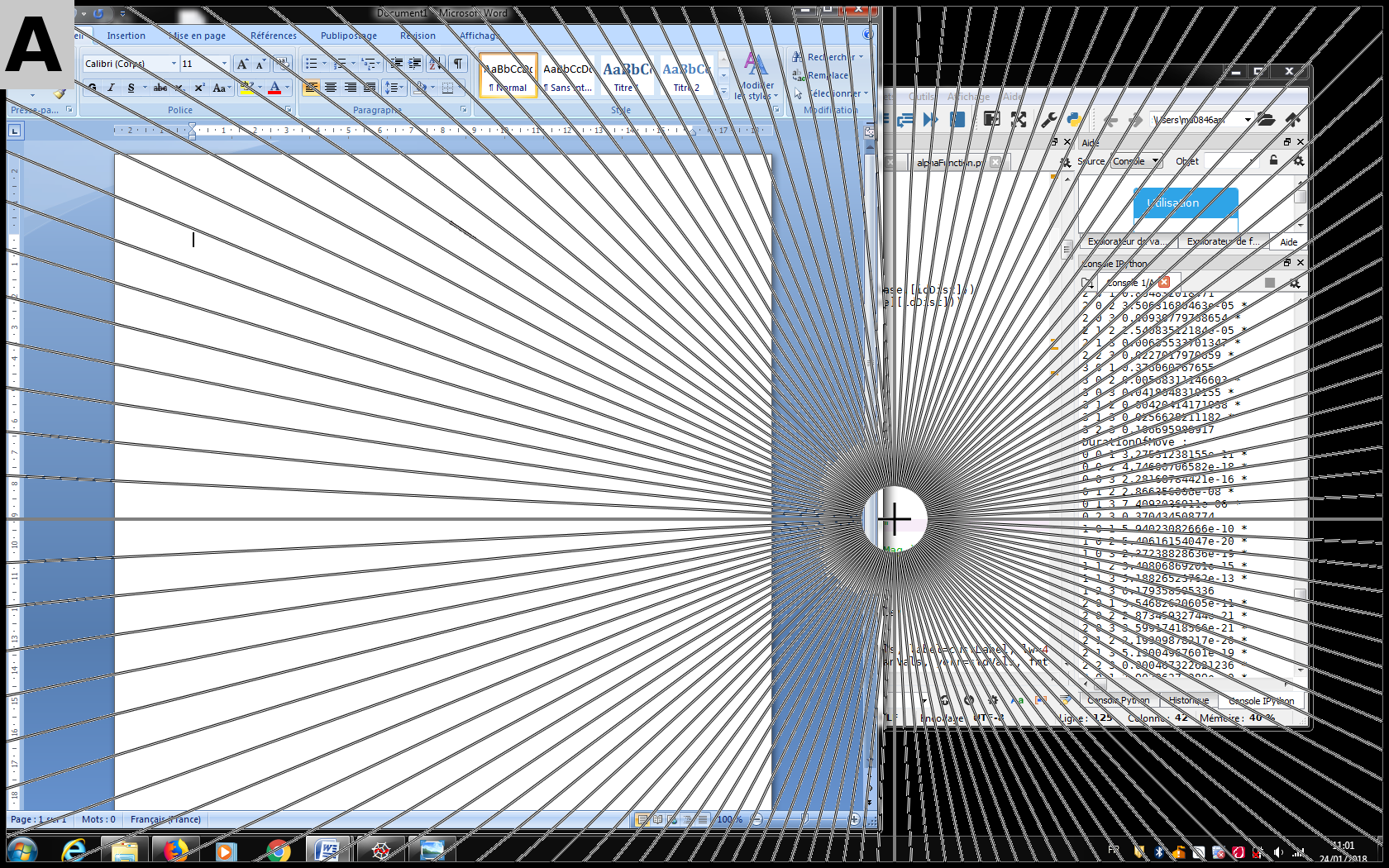} }}%
\qquad
\subfloat{{\includegraphics[width=0.47\linewidth]{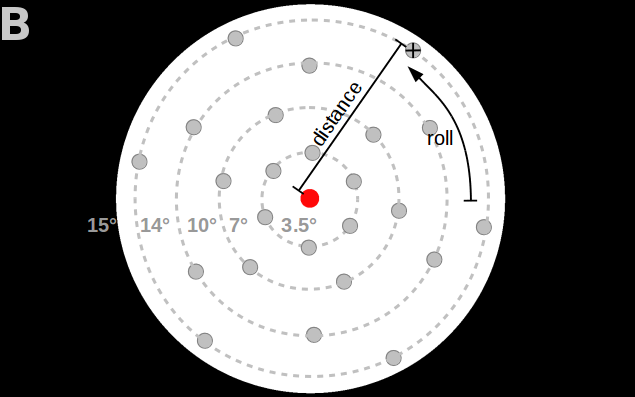} }}%
\caption{\textbf{A}: a screenshot of the Sunny Pointer during its activation over a standard desktop screen. The position of the mouse cursor is materialized by a black cross and the Sunny Pointer displays 128 equidistantly spaced "rays" that radiate from the mouse cursor towards the edges of the screen. \textbf{B}: an annotated scheme of a screen during experiments showing the area of response (white disk), the target (red disk), and the mouse cursor (black cross). Concentric circles and small gray disks are annotations showing the possible positions of the mouse cursor for each trial.}
\label{fig:screenshot_sunnyP}
\end{figure*}

In this work, we tested the capacities of participants to control the trajectory of the mouse cursor under different conditions. The task was to click as quickly as possible on a visual target placed at the center of the computer screen with the mouse cursor initially placed at several random positions. In the following discussion, the distances, positions and sizes of the graphical components displayed on the screen are expressed in a spherical coordinate system with the origin at a point located between the eyes of the participant. The coordinates are angles with regard to the line extending from the origin and reaching the center of the screen. 

The mouse pointer that we developed draws straight lines starting at a short distance from the cursor and covering the screen. The complete $2\pi$ angle around the cursor is divided into several equal portions, each delimited by two lines. In other words, these lines materialize the radiuses of a circle centered on the mouse cursor, thus producing a structure resembling light rays coming from the mouse cursor, which explains the name of the tool. This graphical structure follows each move of the pointer. In order to be easily differentiated from various graphical backgrounds, each line is composed of two colors, one for the inner of the line and one for its borders. In order to limit the parameter space and the complexity of this study, the following setup was chosen for all the participants and all the experiments: 128 lines, each line composed of a black line with a superimposed thinner white line (respectivaly 2 and 1 pixel wide), the lines were constantly displayed when our system was turned on, the lines were starting at a distance from the pointer corresponding to two degrees of visual field when the pointe is at the center of the screen and the length of the lines were ending at the edge of the screen as presented in figure \ref{fig:screenshot_sunnyP}.A.

The visual target to click on was indicated by a 1$^\circ$ wide (diameter) red disk at the center of the screen. The position of the mouse cursor was materialized by a black cross of the same size as the target. The area where the mouse cursor could appear and move was indicated by a gray disk with a diameter equal to the height of the screen and thus corresponding to a radius of 15$^\circ$ of the FOV centered on the screen. The mouse pointer was artificially maintained at the border of the gray area in case the move of the user would have brought it outside. The "rays" of the Sunny Pointer were completely hidden outside this gray area. For each exercise, four initial pointer-target distances were used: 3.5$^\circ$, 7$^\circ$, 10.5$^\circ$ and 14$^\circ$. For each pointer-target distance, six pointer-target roll angles were used. For a given distance, the first of the six roll angles was uniformly chosen over a range of [$0-2\pi$] and the five other angles were equidistantly spaced with $\pi/3$ starting from the first randomly chosen angle.  An annotated scheme of the screen during experiments is shown in figure \ref{fig:screenshot_sunnyP}.B.

In each exercise, participants thus had to perform 6 trials for each of the four distances, making a total of 6x4=24 trials for each exercise. All the trials in each exercise were randomly permuted. Before each trial, the participant had to replace the mouse device at the center of the assigned moving area on the table. When the user was ready, he or she had to press the left mouse button. Immediately afterward, the mouse cursor was placed in a new position and the participant had to move it in order to click on the target with the left button as quickly as possible. When the target was clicked on, the pause to replace the mouse device at the center of the designated area was repeated before the next trial. Before the experiment, a questionnaire recorded the age, gender, laterality, vision disorders and acuity, and a subjective evaluation of the ability of the participant to use the mouse. A verbal explanation of the different tasks was briefly presented to all subjects before the start of the experiment.

The experimental apparatus consisted of a PC running Windows 7. The software was programmed using C\#. A chinrest was placed in front of the computer screen so that the user’s eyes were horizontally in line with the center of the screen. The screen was placed so that the angle between this horizontal axis and the top edge of the screen corresponded to 15$^\circ$ of the participant's visual field. For example, a screen 52cm wide and 32cm high should be placed approximately 59.7cm away from the chinrest. The mouse was placed on the table to the right of the chair (since only right handed subjects took part in the experiment) in the middle of an area large enough for the user to move the mouse freely. A logitech B110 optical mouse featuring a sensitivity of 800 dpi was used. In the operating system, to ensure that each participant had to do the same physical arm displacement accross the trials, sensitivity to the acceleration of the mouse was disabled and the sensitivity of the operating system mouse was set in such a way that the height of the screen could be crossed with a move of the mouse device of 3.5 cm on the table. For example, the height of a screen with a resolution of 1680x1050 requires a sensitivity of approximately 760dpi to be entirely vertically crossed in 3.5cm. The program sampled and recorded the mouse position (X and Y) at a rate of 33 Hz. There was no perceptible lag between movement of the mouse device and the associated movement of the cursor. 

\subsection{General Description of the Analysis}
\label{analysis_descr}

In the following work, the time to complete the task (TCT) is considered the time lapse between the moment the target and the pointer are displayed on the screen and the moment the participant clicks on the target with the pointer. As proposed in \cite{Card_1980}, we distinguish three periods in the TCT: the acquisition time (AT) is the time lapse between the moment the target and pointer have been displayed on the screen and the participant’s first move. This first move is detected when the mouse cursor has been moved by more than 10 pixels from its initial position. This is the time used by the participant to collect information and to plan his move. The movement time (MT) is the period starting with the first detected move of the pointer and ending when the participant reaches the target for the last time (in case of multiple attempts due to overshoots). This is thus the theoretical moment at which the participant could have validated the trial if no time was required to click on the mouse button. The keystroke time (KT) is the period starting when the pointer reaches the target for the last time until the user clicks on it and thus completes the task. 

In order to compare the results of one configuration to those of another, we use the Mann–Whitney U test to determine the probability that the two sets of results come from the same distribution. This test does not assume a normal distribution of the samples and can be applied to two non-paired and independent samples.

\section{Experiment 1}

\label{exp1_explanation}

\begin{figure*}[ht]\centering
\subfloat{{\includegraphics[width=0.47\linewidth]{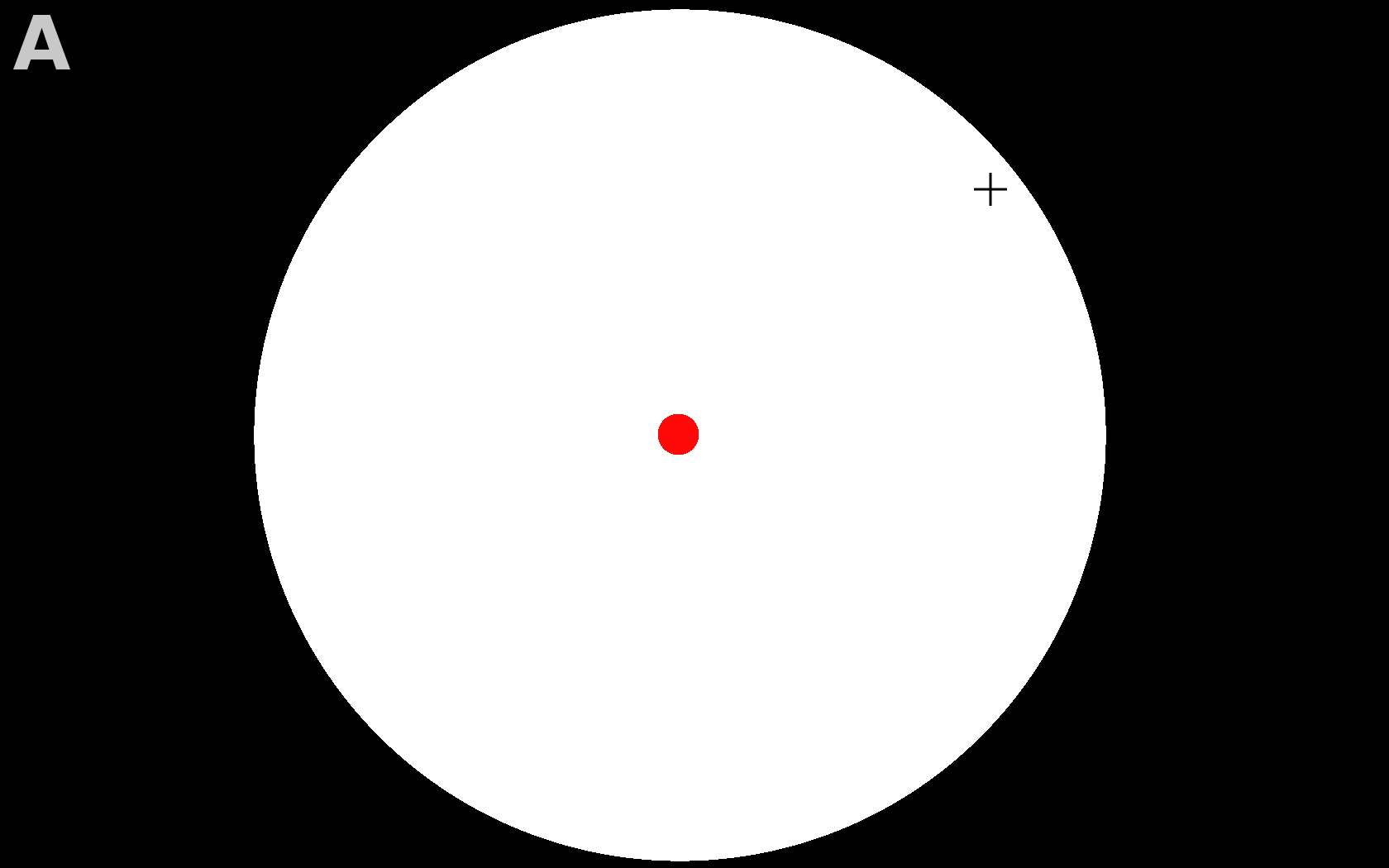} }}%
\qquad
\subfloat{{\includegraphics[width=0.47\linewidth]{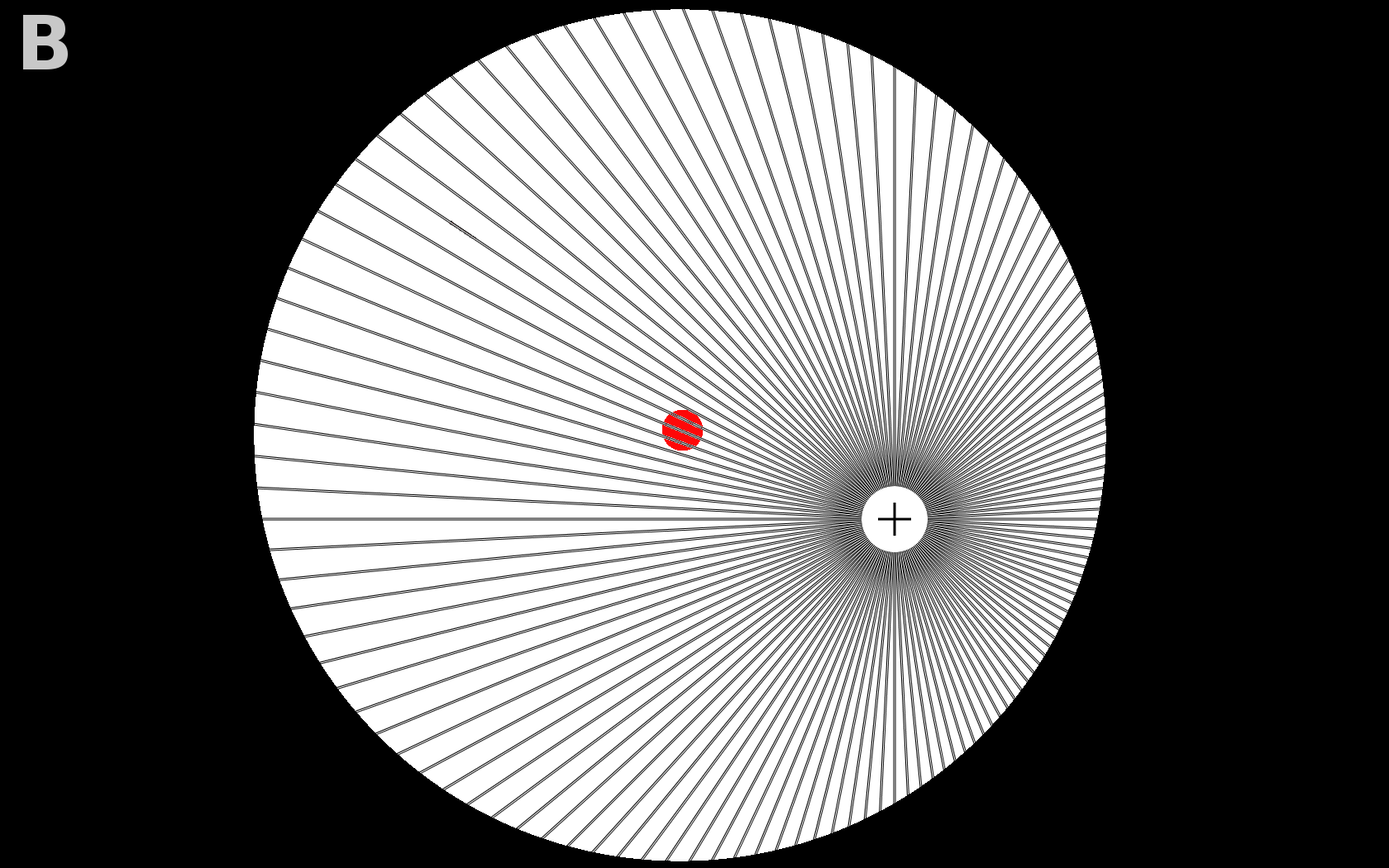} }}%
\caption{\textbf{A}: a screenshot when the Sunny Pointer is turned off, i.e. the CP-PVL condition (Crosshair Pointer - Peripheral Vision Loss). \textbf{B}: a screenshot when the Sunny Pointer is turned on, i.e. the SP-PVL condition (Sunny Pointer - Peripheral Vision Loss).}
\label{fig:illustr_exo1}
\end{figure*}

In this experiment, we measured the improvement due to the Sunny Pointer compared to a regular use of the mouse with six participants with different types of Peripheral Vision Loss summarized in the table \ref{tab:participant}. 

\begin{table}\centering
\begin{tabular}{|c|c|c|c|}
        \toprule
        Participant & Bino. VF  & Bino. Acc. & Vis. Dis. \\
                & radius (deg.) & (10\textsuperscript{e} Parinaud) & \\

        \midrule
        A & $1.25$ & $7$ & RP \\
        B & $2$ & $4.5$ & RP \\
        C & $2.5$ & $7$ & GL \\
        D & $2.5$ & $2$ & GL \\
        E & $2.5$ & $2$ & RP \\
        F & $5$ & $1.6$ & RP \\
        \bottomrule
    \end{tabular}
\vspace{0.3cm}
\caption{Table summarizing the participant id (first col.), the corresponding radius of the binocular visual field express in degree (second col.), the corresponding binocular acuity given in 10\textsuperscript{e} in the test of Parinaud (third col.), and the type of visual disorder (RP: Retinitis Pigmentosa, GL: Glaucoma).}
\label{tab:participant}
\end{table}

The experiment was composed of 5 successive exercises, each composed of 24 trials as explained in the section \ref{exp_descr}. We used two types of exercises as illustrated in figures \ref{fig:illustr_exo1}.A and \ref{fig:illustr_exo1}.B. In the first type of exercise the Sunny Pointer is off and the pointer is only materialized by a small black crosshair as shown in the left figure. This type of exercise is referred as the CP-PVL condition (Crosshair Pointer - Peripheral Vision Loss). The second type of exercise, shown in the figure on the right, is referred to as the SP-PVL condition (Sunny Pointer - Peripheral Vision Loss). The conditions of this exercise are the same as the previous one except that our pointer is turned on, displaying lines radiating from the mouse cursor as depicted in section \ref{exp_descr}. Each participant passed a first exercise in the CP-PVL condition, then 3 exercises in the SP-PVL condition and then again one exercise in the CP-PVL condition. The results presented in the figure \ref{fig:resultats_exp_all} are aggregated results of the exercises of each type.

\subsection{Results of the experiment 1}

\begin{figure*}[ht]\centering
\subfloat{{\includegraphics[width=0.47\linewidth]{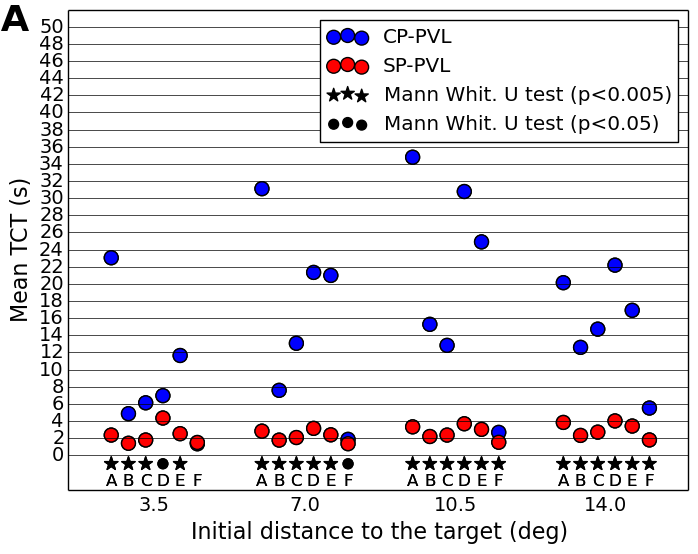} }}%
\qquad
\subfloat{{\includegraphics[width=0.47\linewidth]{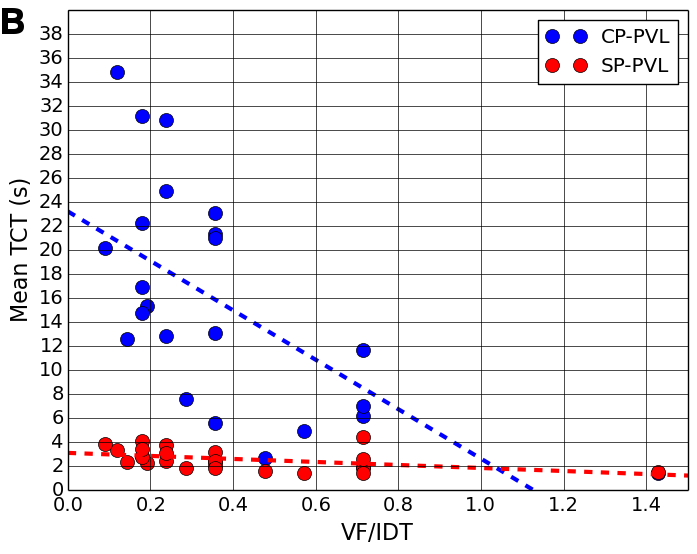} }}%
\caption{\textbf{A}: Mean Time to Complete the Task (TCT) without (CP-PVL) and with our pointer (SP-PVL) as a function of the initial distance from the pointer to the target for each of the six participants [A,B,C,D,E,F]. \textbf{B}: Mean Time to Complete the Task (TCT) without (CP-PVL) and with our pointer (SP-PVL) as a function of the ratio between the radius of the visual field of the user (VF) and the initial distance from the pointer to the target (IDT) for all the participants. Linear regressions are superimposed with dashed lines.}
\label{fig:resultats_exp_all}
\end{figure*}

The comparison between the mean Time to Complete the task (TCT) without (CP-PVL) and with our pointer (SP-PVL) as a function of the initial distance to the target is shown for each of the six participants in the figure \ref{fig:resultats_exp_all}.A. While using our pointer the TCT ranges from 1 to 4 seconds (mean=2.5s), the TCT in the CP-PVL condition is more variable with a mean value of 17 seconds with results that highly depend both on the participant and on the initial distance. All the comparisons between CP-PVL and SP-PVL were significants except for the participant F with an initial pointer distance of 3.5°.

The high variability in the CP-PVL condition is due to the visual field of the participant and the strategies they used. As they told us after the exercises, some of the participant were doing an horizontal scanning from top to bottom in order to find the mouse pointer, some other were doing a scanning in spiral either from the outside toward the center or the inverse, and some other do not seem to use a specific strategy. Concerning the participant F, its visual field of 5° is large enough to simultaneously see the target and the pointer for an initial distance of 3.5° but the TCT differences become more and more significant as the initial distance to the target exceed its visual field. 

To better understand the relation between the improvement due to the use of the Sunny Pointer and the conditions of its use, we plotted in the figure \ref{fig:resultats_exp_all}.B the mean TCT as a function of the ratio between the radius of the Visual Field of the user (VF) and the initial distance from the pointer to the target (IDT). Results for all the participants in CP-PVL and the SP-PVL conditions are shown superimposed by their respective linear regression in dashed lines. Not surprisingly, the advantage of our system strongly depends on the ratio between the visual field and the initial distance to the target. The improvement seems to appear for a ratio lower than 1 and reaches a value of 7 for a ratio close to 0.1. The coefficients of determination of the linear regressions in the CP-PVL condition and in the SP-PVL condition are respectively 0.41 and 0.15. We found no relation between the time to complete the task and the accuity.

\subsection{Conclusion of the experiment 1}

The sunny pointer can decrease the time to click on a target by a large factor compare to the standard use of the mouse. This seems to be valid for peripheral vision losses due to both Retinitis Pigmentosa and Glaucoma. However, it seems that people use the system in various ways. Some of the people were telling us during the debriefing that they were using the lines in order to find the pointer quickly and afterwards moving the pointer as they would have done without Sunny Pointer. Others were directly starting the move towards the target just based on the convergence of the lines that were covering it. We precisely describe this last strategy in the next experiment.

\section{Experiment 2}

Jean (the name has been voluntarily changed) is 36 years old. Due to a retinitis pigmentosa diagnosed 15 years ago, he has a field of view restricted to approximately 3.5$^\circ$ around the direction of his gaze.  His remaining central acuity is 3/10 and 4/10 but reaches 9/10 and 10/10 with the correction of the glasses worn during this experiment. He was a computer developer before his impairment and is thus used to handling the mouse. Before this experiment, he was mainly placing the mouse pointer to the top left corner of the screen after each click of the mouse and, after having found a new target, he was moving it towards the estimated new target direction while following the pointer with the eyes on the screen.

In order to understand how someone with PVL can use our tool after some training, the Sunny Pointer was installed on his computer 2 weeks before the experiment took place and we asked him to use it every day. He used the system on average one hour per day. Two weeks later, we used an eye tracker (Tobii Pro TX300) during task performance to record the participant's gaze direction at a rate of 33 Hz. We did not experience any calibrating issues despite the wearing of glasses. The gaze was classified into three categories according to the following procedure: if the gaze is situated in a radius of less than 2$^\circ$ from the target, the gaze is categorized as being focused on the target; if not, and if the gaze is situated in a radius of less than 2$^\circ$ from the current mouse cursor position, the gaze is categorized as being focused on the mouse cursor; the gaze is categorized as being focused elsewhere if the two above conditions are not satisfied. For each trial, the gaze direction profile was renormalized on a time scale between 0 and 1, in which 0 is the time at which the cursor is displayed on the screen and 1 is the time at which the participant clicks on the target.

\subsection{Results of the experiment 2}
\label{exp1_results}

\begin{figure*}[ht]\centering
\subfloat{{\includegraphics[width=0.47\linewidth]{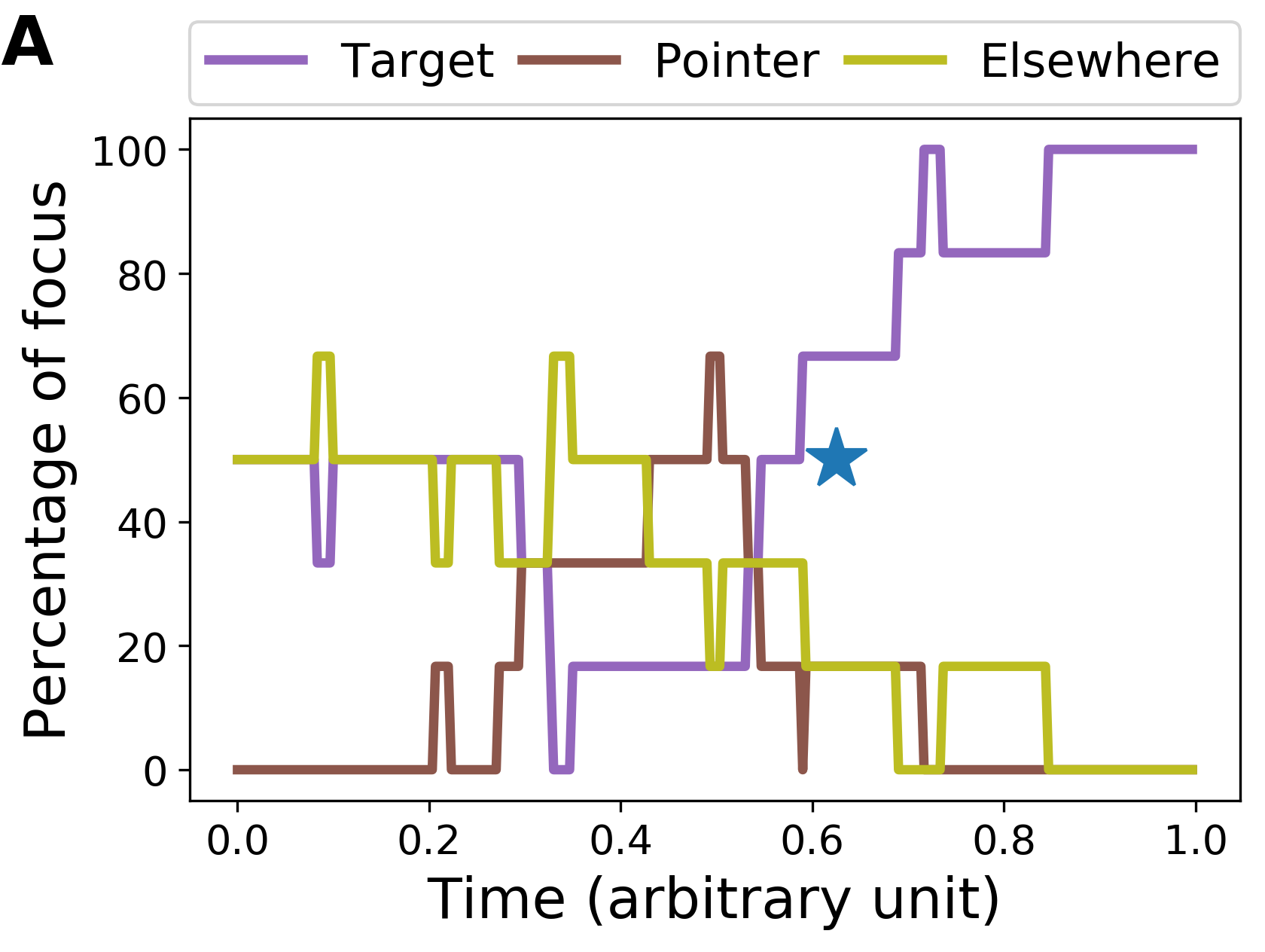} }}%
\qquad
\subfloat{{\includegraphics[width=0.47\linewidth]{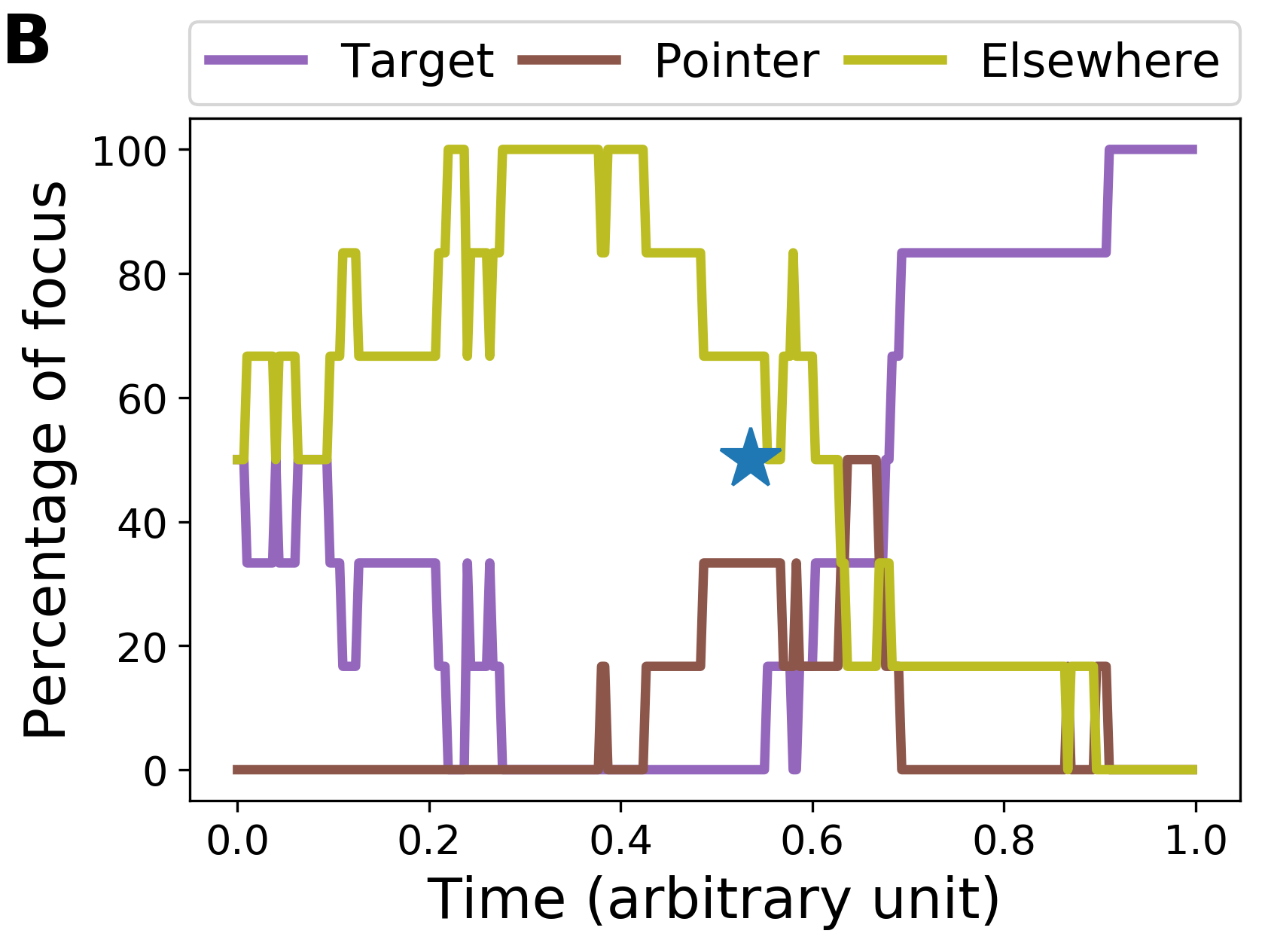} }}%
\qquad
\subfloat{{\includegraphics[width=0.47\linewidth]{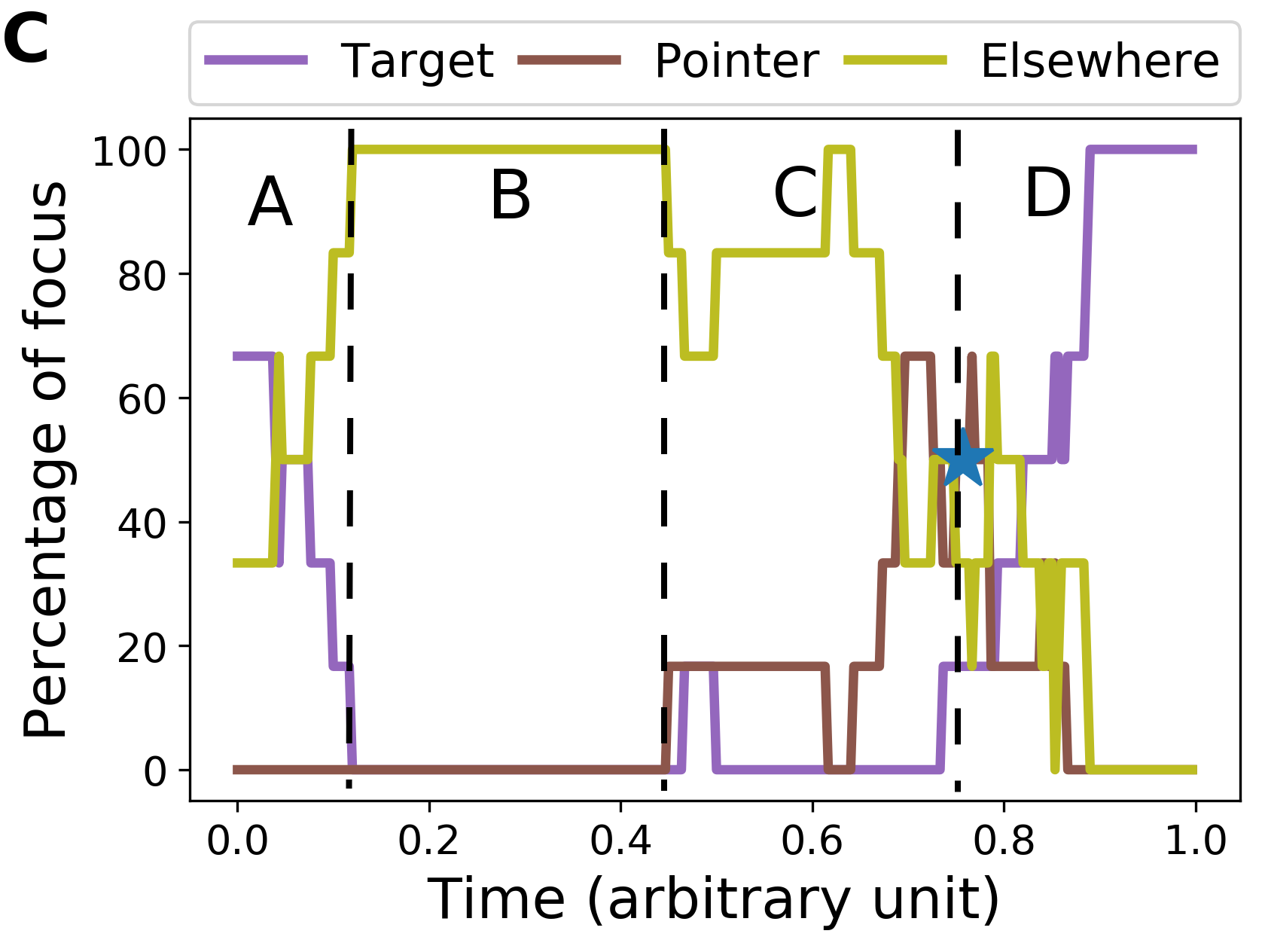} }}%
\qquad
\subfloat{{\includegraphics[width=0.47\linewidth]{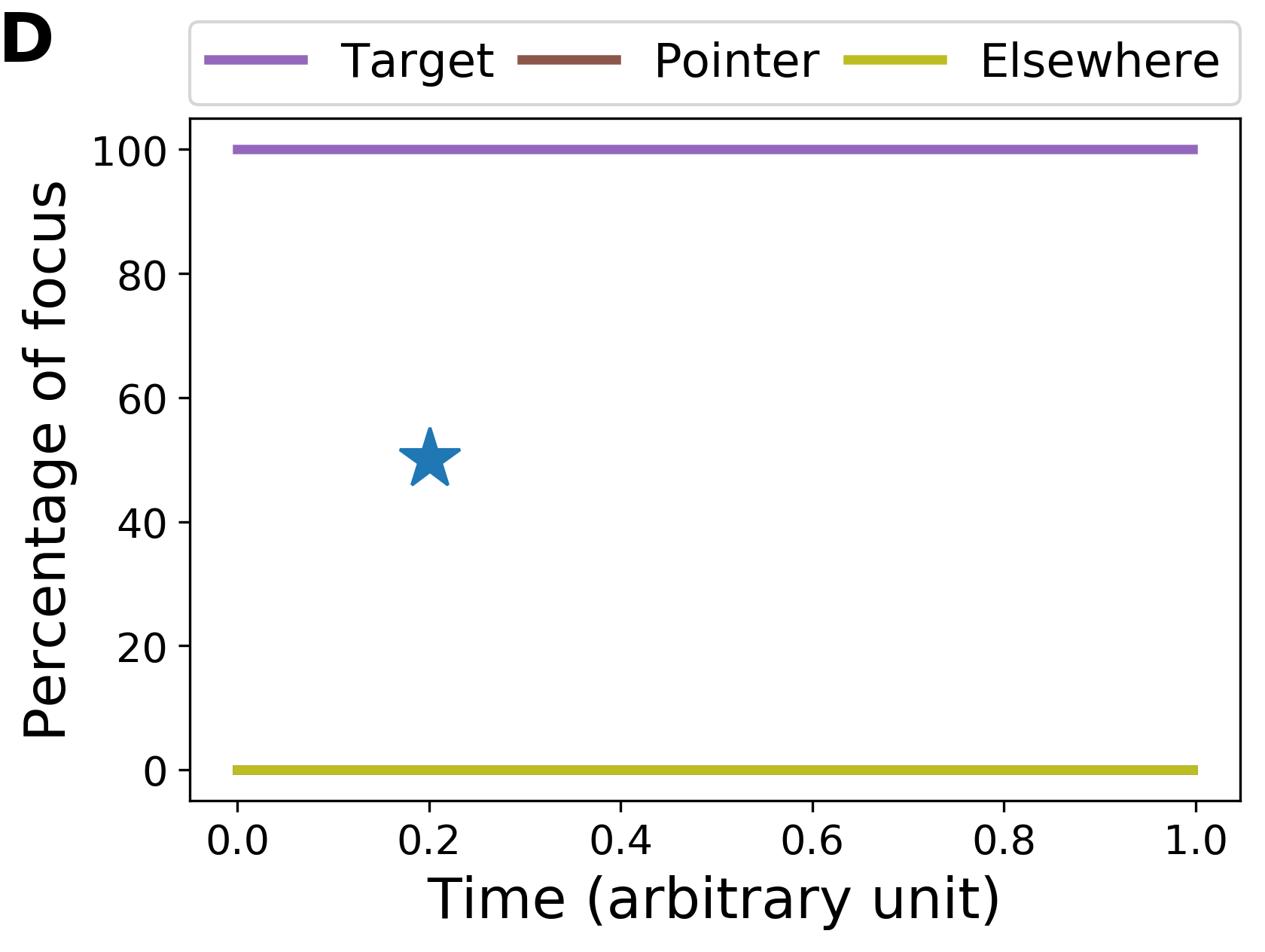} }}%
\caption{Figures showing the proportions of time the participant focused on the target, on the mouse cursor, or elsewhere on the screen. The blue star materializes the mean time of the first detected move. \textbf{A}: gaze recordings for an initial pointer-target distance of 3.5$^\circ$ in the CP-PVL condition (Crosshair Pointer - Peripheral Vision Loss). \textbf{B}: gaze recordings for an initial pointer-target distance of 7$^\circ$ in the CP-PVL condition. \textbf{C}: gaze recordings for an initial pointer-target distance of 14$^\circ$ in the CP-PVL condition. \textbf{D}: gaze recordings for an initial pointer-target distance of 14$^\circ$ in the SP-PVL condition.}
\label{fig:results_eye_tracking}
\end{figure*}

As for the previous experiment, we used 2 types of exercises: one in the CP-PVL condition (Sunny Pointer turned off) and the other in the SP-PVL condition. The averaged gaze direction profiles in the CP-PVL condition  for 3.5$^\circ$, 7$^\circ$ and 14$^\circ$ initial pointer-target distances are shown in figures \ref{fig:results_eye_tracking}.A, \ref{fig:results_eye_tracking}.B, and \ref{fig:results_eye_tracking}.C respectively. Data for 10.5$^\circ$ in the CP-PVL condition are not shown as they conform to the general scheme presented in the previous 3 plots. The averaged gaze direction profiles in the SP-PVL condition (Sunny Pointer turned on) for an initial pointer-target distance of 14$^\circ$ are shown in figure \ref{fig:results_eye_tracking}.D. Data in the SP-PVL condition at 3.5$^\circ$, 7$^\circ$ and 10$^\circ$ are not shown since they reproduce the data for 14$^\circ$, i.e. a constant focus on the target. 

As annotated in figure \ref{fig:results_eye_tracking}.C, the strategy used in the CP-PVL condition can be broken down into a sequence of 4 steps: (A) a proximity search in which the participant looks in the vicinity of the target to find the cursor; (B) a further search on the complete screen if the cursor was not found during the proximity search; (C) the localization of the pointer and the estimation of the proper direction of movement; and (D) the moving of the pointer towards the target. The relative duration of the steps depends on the initial distance between the mouse cursor and the target. The longer the pointer-target distance, the longer the search on the screen. The participant explained this strategy as follows "I first seek the mouse cursor by following a spiral centered on the target. Sometimes, I also try my luck doing a random search. When I have found it,  I move the cursor in the estimated direction of the target until both the cursor and the target enter my field of view."

On the contrary, the profile of the gaze when the Sunny Pointer is turned on is completely different, as presented in figure \ref{fig:results_eye_tracking}.D. The participant focuses exclusively on the target during the whole trial. This confirms the strategy that he explained in these terms after the experiment: "I constantly look at the target and I use the direction and the convergence of the lines to evaluate the direction in which I have to move the mouse." This strategy is similar to the first one depicted in \cite{Smith_2000} concerning full visual field users. This illustrates the first main advantage of our pointer: it not only facilitates the localization of the mouse pointer but also eliminates the need to precisely locate the pointer in the 2-dimensional space of the screen before starting the mouse move.

\begin{figure*}[ht]\centering
\subfloat{{\includegraphics[width=0.47\linewidth]{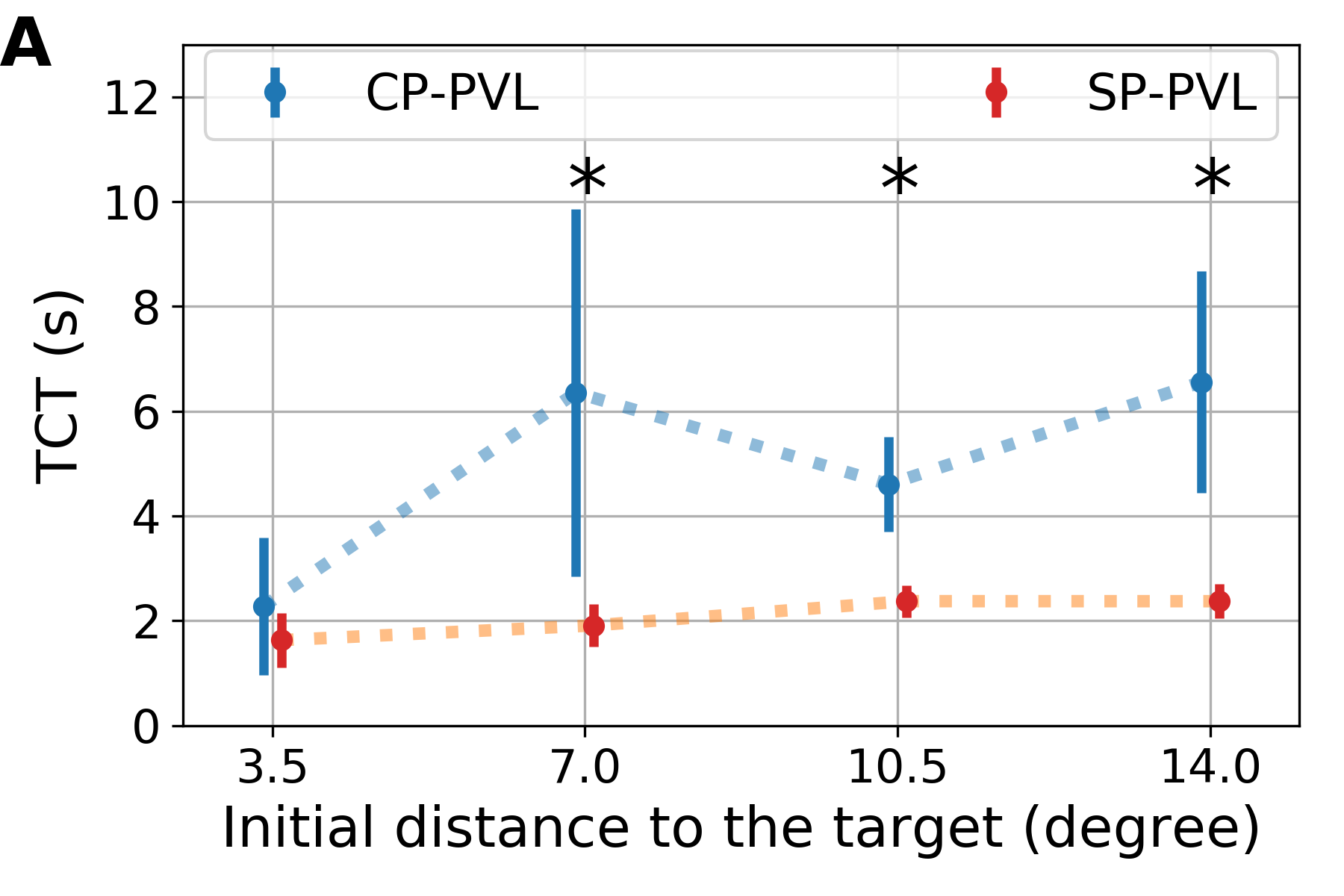} }}%
\qquad
\subfloat{{\includegraphics[width=0.47\linewidth]{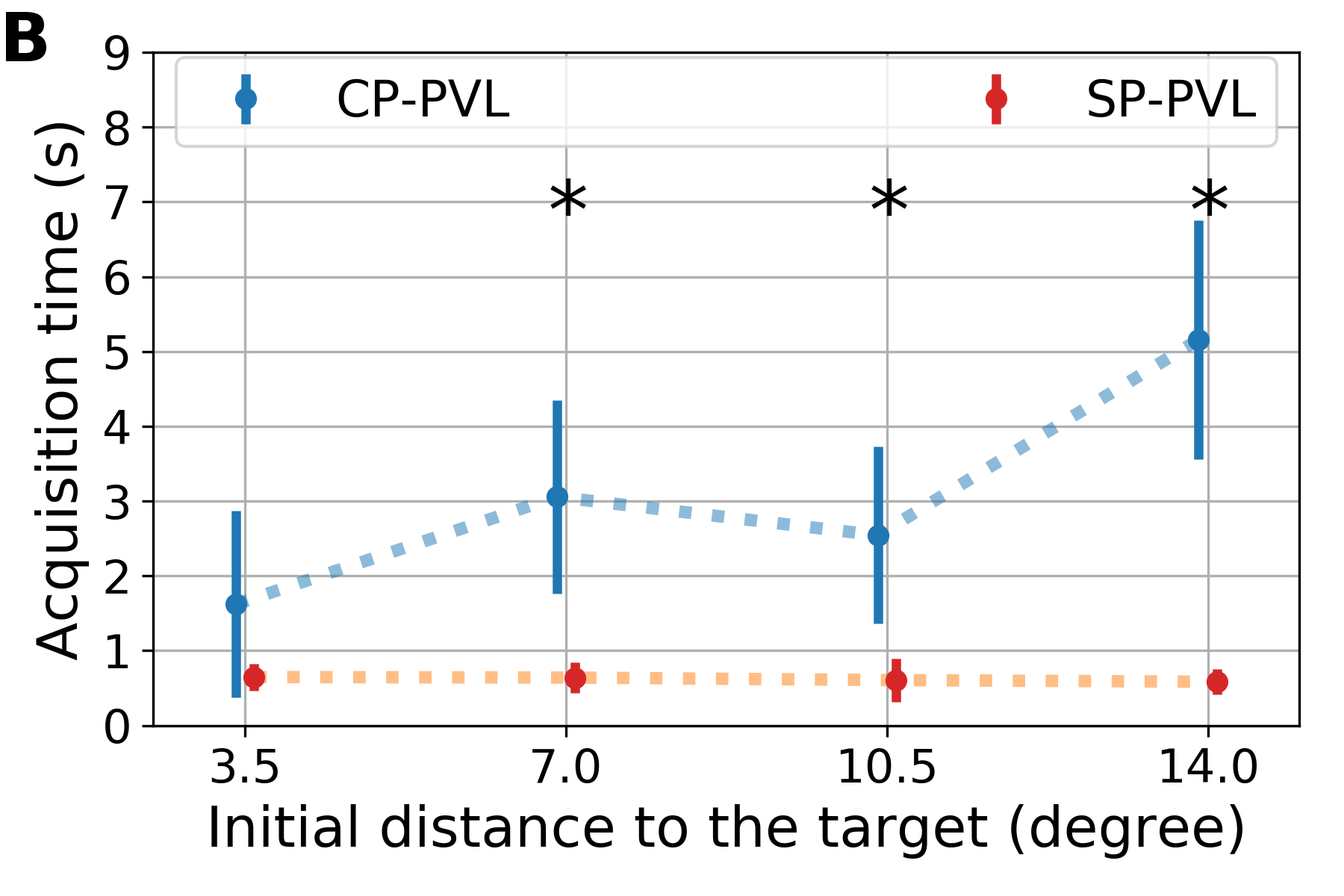} }}%
\qquad
\subfloat{{\includegraphics[width=0.47\linewidth]{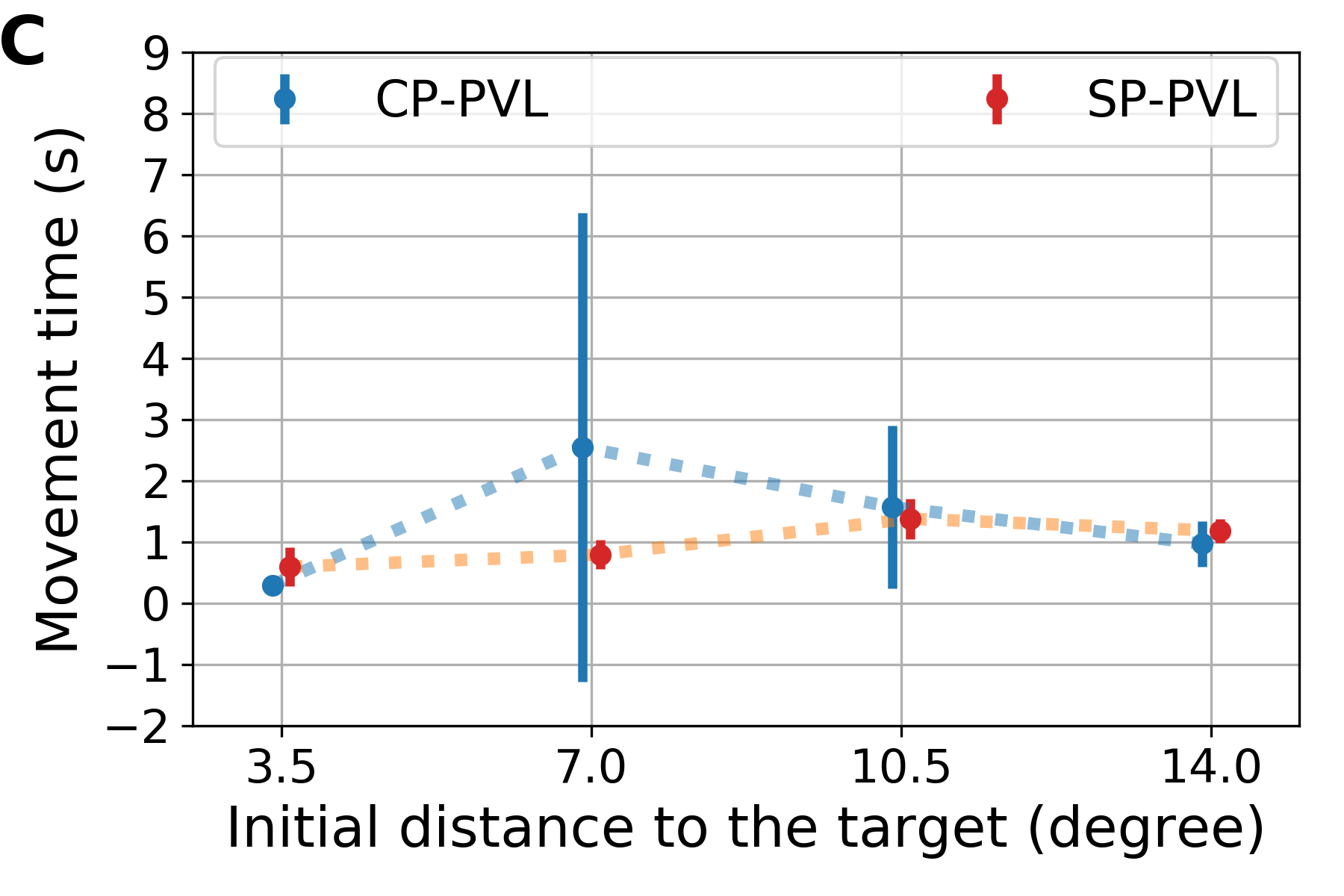} }}%
\qquad
\subfloat{{\includegraphics[width=0.47\linewidth]{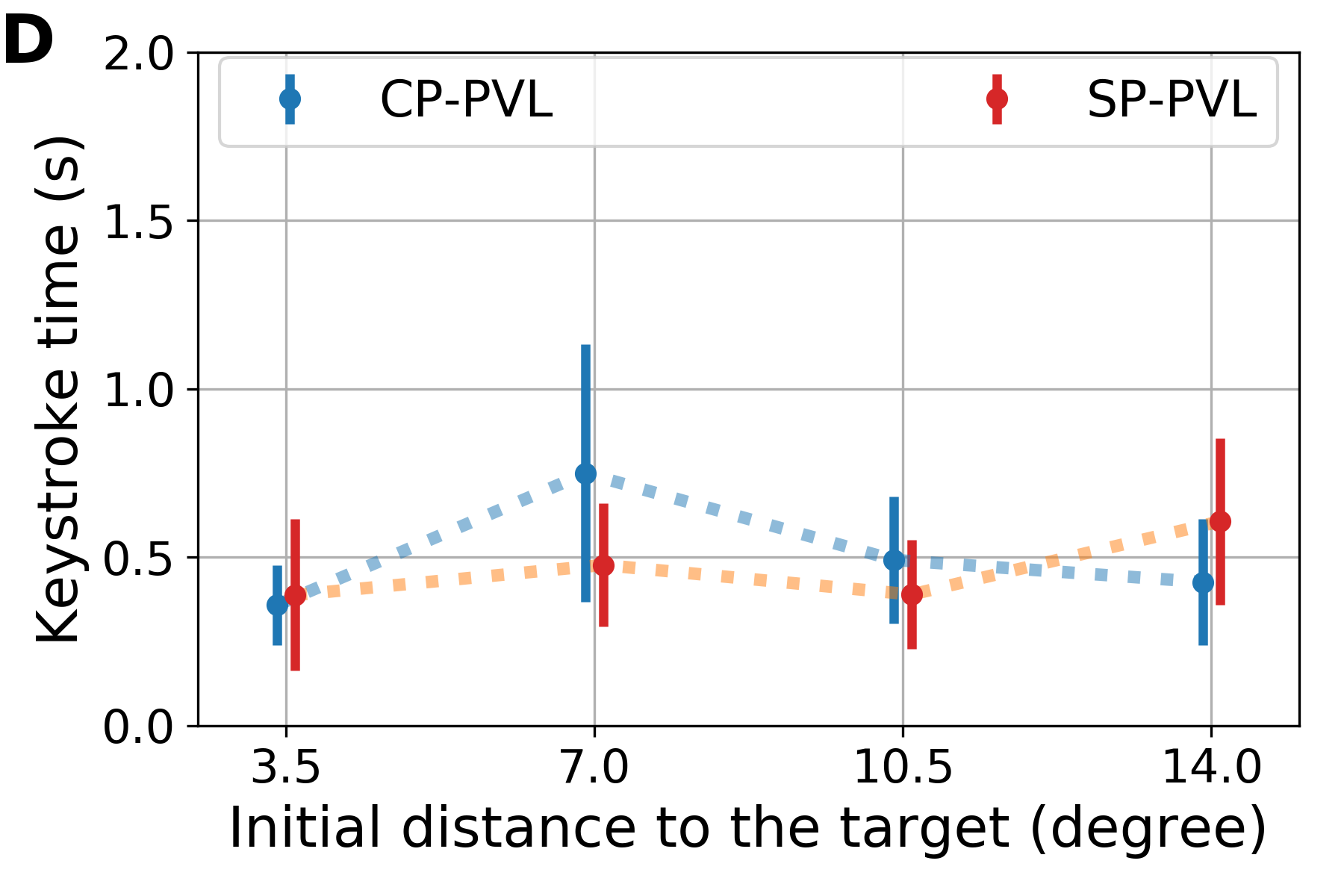} }}%

\caption{\textbf{A}: the time to complete the task (TCT) against the initial distance between the cursor and the target in the CP-PVL (Crosshair Pointer - Peripheral Vision Loss) condition (blue) and the SP-PVL (Sunny Pointer - Peripheral Vision Loss) condition (red). \textbf{B}: the acquisition time (AT). \textbf{C}: the movement time (MT). \textbf{D}: the time to click on the target (keystroke time: KT). Dotted lines are superimposed to show the variations in mean times. Stars materialize conditions in which the distributions in CP-PVL and SP-PVL conditions significantly differ using a Mann-Whitney U test ($p<0.05$).} 
\label{fig:results_exp1_complete}
\end{figure*}

As shown in figure \ref{fig:results_exp1_complete}.A, this drastic change in strategy reduces the time to complete the task (TCT) by a factor of up to 3, except in the case of the 3.5$^\circ$ distance. In this last configuration, Jean's VF of $\pm3.5^\circ$ was sufficient to quickly localize the target and the pointer and the assistance afforded by our pointer was thus not as significant as in the case of longer pointer-target distances. It should be noted that the standard deviations in the SP-PVL condition for 7$^\circ$, 10.5$^\circ$ and 14$^\circ$ distances are 10 times lower than those observed in the CP-PVL condition, showing that the information provided by the Sunny Pointer is reliable enough to induce a reproducible TCT. 

As shown in figure \ref{fig:results_exp1_complete}.B, most of the decrease observed in the TCT results from an important drop in the acquisition time in the SP-PVL compared to the CP-PVL condition (see section \ref{analysis_descr} for a definition of the acquisition time). In the CP-PVL condition, the mean acquisition time ranges from 1.5 seconds for a pointer-target distance of 3.5$^\circ$ up to 5 seconds for larger pointer-target distances. The use of the Sunny Pointer, i.e. the results obtained in the SP-PVL condition, reduces this time to approximately half a second, independent of the initial pointer-target distance. This decrease in acquisition time is the first advantage of the Sunny Pointer. 

Figure \ref{fig:results_exp1_complete}.C shows that in the SP-PVL condition, the movement time slowly increases from approximately half a second to a little more than one second. The situation when our pointer is off (CP-PVL condition) is more complicated and the mean movement time does not significantly differ from the SP-PVL condition except in the 7$^\circ$ condition. During exchanges with the participant after the experiment, he mentioned trials in the CP-PVL condition in which he became confused about the position of the mouse cursor during its move towards the target. He thus stopped the move until he once again localized the mouse pointer before completing the trial. The Sunny Pointer prevents these confusions from happening, which is the second advantage of its use. 

Keystroke times shown in figure \ref{fig:results_exp1_complete}.D do not significantly differ from SP-PVL to CP-PVL conditions, showing an average value of approximately $0.5s$.

\subsection{Conclusion of the experiment 2}

The Sunny Pointer can drastically change the way people with PVL use the computer mouse. Without the Sunny Pointer, the user searches the screen for the mouse cursor before starting to move it in the estimated direction of the target. With the help of the Sunny Pointer, the user can focus solely on the target and uses the information provided by the "rays" of the pointer to determine how to best move the mouse cursor. The 2-dimensional localization of the pointer on the screen is no longer necessary and this brings with it a significant reduction in acquisition time, resulting in a decrease of the TCT by a factor 3. Moreover, the use of the Sunny Pointer seems to decrease the probability of confusions during the pointer move towards the target.

\section{Experiment 3}

\begin{figure*}[ht]\centering
\subfloat{{\includegraphics[width=0.47\linewidth]{./figures_letters/exo_1.png} }}%
\qquad
\subfloat{{\includegraphics[width=0.47\linewidth]{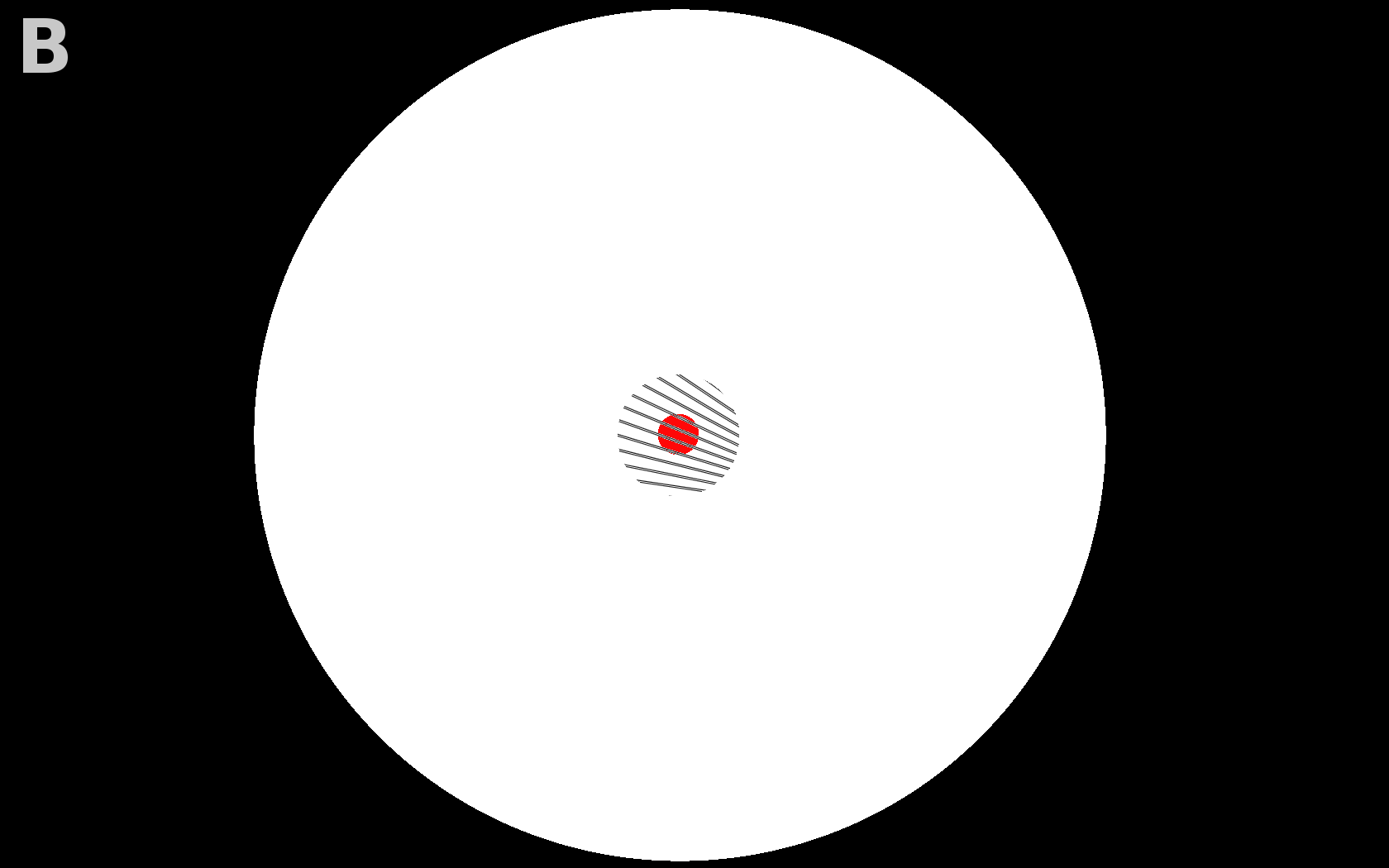} }}%
\caption{\textbf{A}: a screenshot in the CP-FVF condition (Crosshair Pointer - Full Visual Field). \textbf{B}: a screenshot of the SP-SIMPVL (Sunny Pointer - Simulated Peripheral Vision Loss) condition (radius 1.5$^\circ$) and the Sunny Pointer turned on.}
\label{fig:illustr_exp_2}
\end{figure*}

In the third experiment, we tried to understand how to improve the performances obtained with the Sunny Pointer in order to bring them closer to those obtained by participants with full visual fields (FVF) using a standard mouse pointer. To this end, we studied the targeting performances of FVF participants with and without a simulated PVL. Twenty participants aged from 15 to 35 years old participated in this third experiment. All participants were right handed with a corrected visual acuity equal to or greater than 8/10 in the worst eye. They had no motor disabilities and subjectively evaluated their ability to move the mouse at 7/10 or higher. 

This experiment was composed of 4 exercises and was organized as follows. The first two exercises were identical to the two described in the first experiment (section \ref{exp1_explanation}). The first exercise thus consisted in a normal use of a standard mouse pointer (a black cross) and a target materialized by a red disk (see figure \ref{fig:illustr_exp_2}.A for an illustration). This exercise is referred to as the CP-FVF condition (Crosshair Pointer - Full Visual Field). The second exercise was identical to the SP-PVL condition in the first experiment. But unlike the first experiment, results obtained during this exercise were not analyzed. It was simply used as a preliminary exercise to accustom participants to the Sunny Pointer. In the third and fourth exercises, a mask completely hiding the screen except for a round aperture of 1.5$^\circ$ radius placed at the target position (the center of the screen) was superimposed in order to force the subjects to use only their central vision, as presented in figure \ref{fig:illustr_exp_2}.B. In other words, the mouse cursor and the "rays" of the pointer were hidden except for a small area around the target. This is thus similar to the "Window" paradigm used in \cite{larson2009}. This paradigm simulates a PVL impairment to the extent that participants can only use the lines of the Sunny Pointer that are visible in a restricted area to move the mouse cursor towards the target. The third exercise is designed to get the participants used to the simulated PVL and the fourth exercise is referred to as the SP-SIMPVL condition (Sunny Pointer - Simulated PVL) whose results are compared to those of the CP-FVF condition in the following figures.

\subsection{Results of the experiment 3}

\begin{figure*}[ht]\centering
\subfloat{{\includegraphics[width=0.47\linewidth]{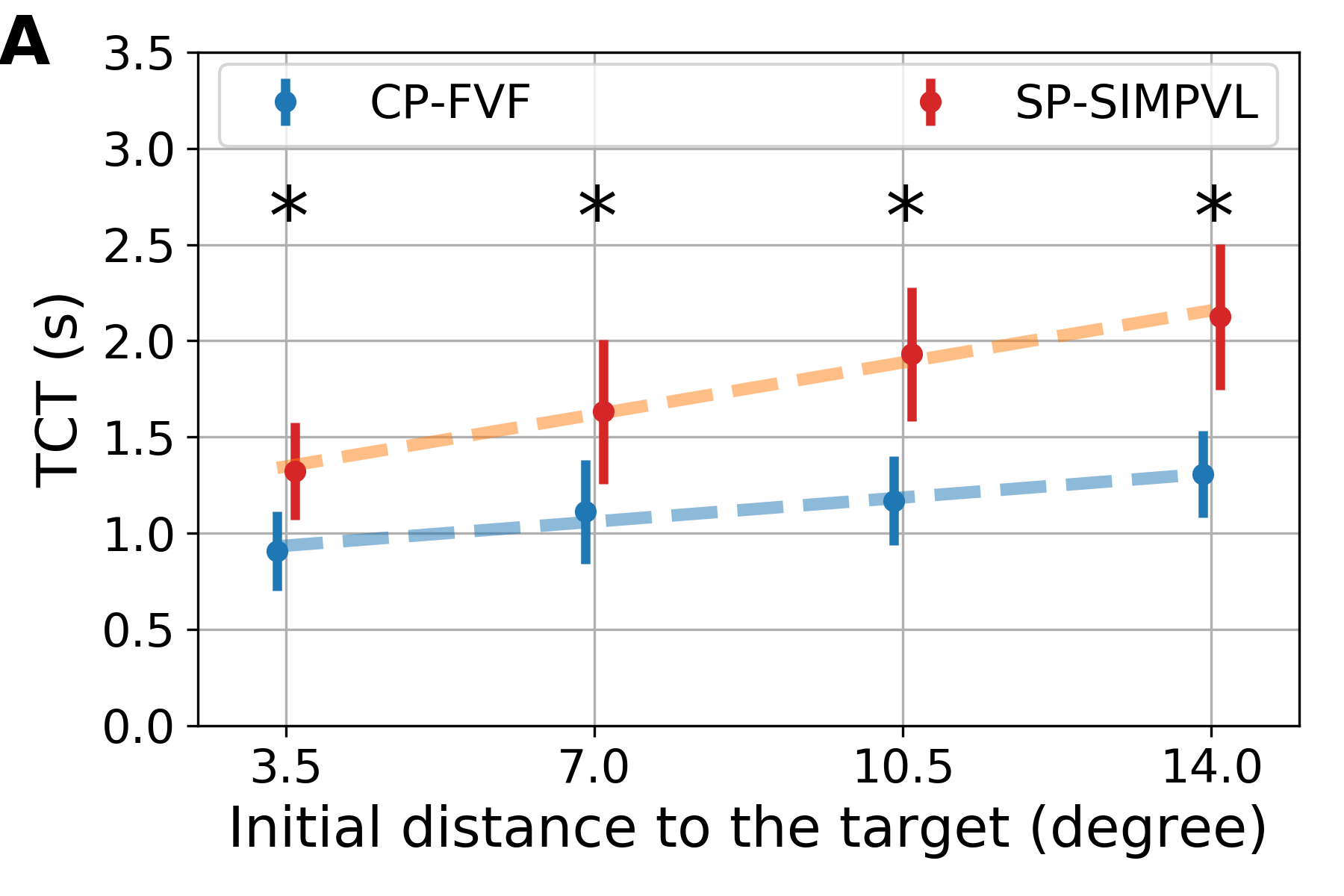} }}%
\qquad
\subfloat{{\includegraphics[width=0.47\linewidth]{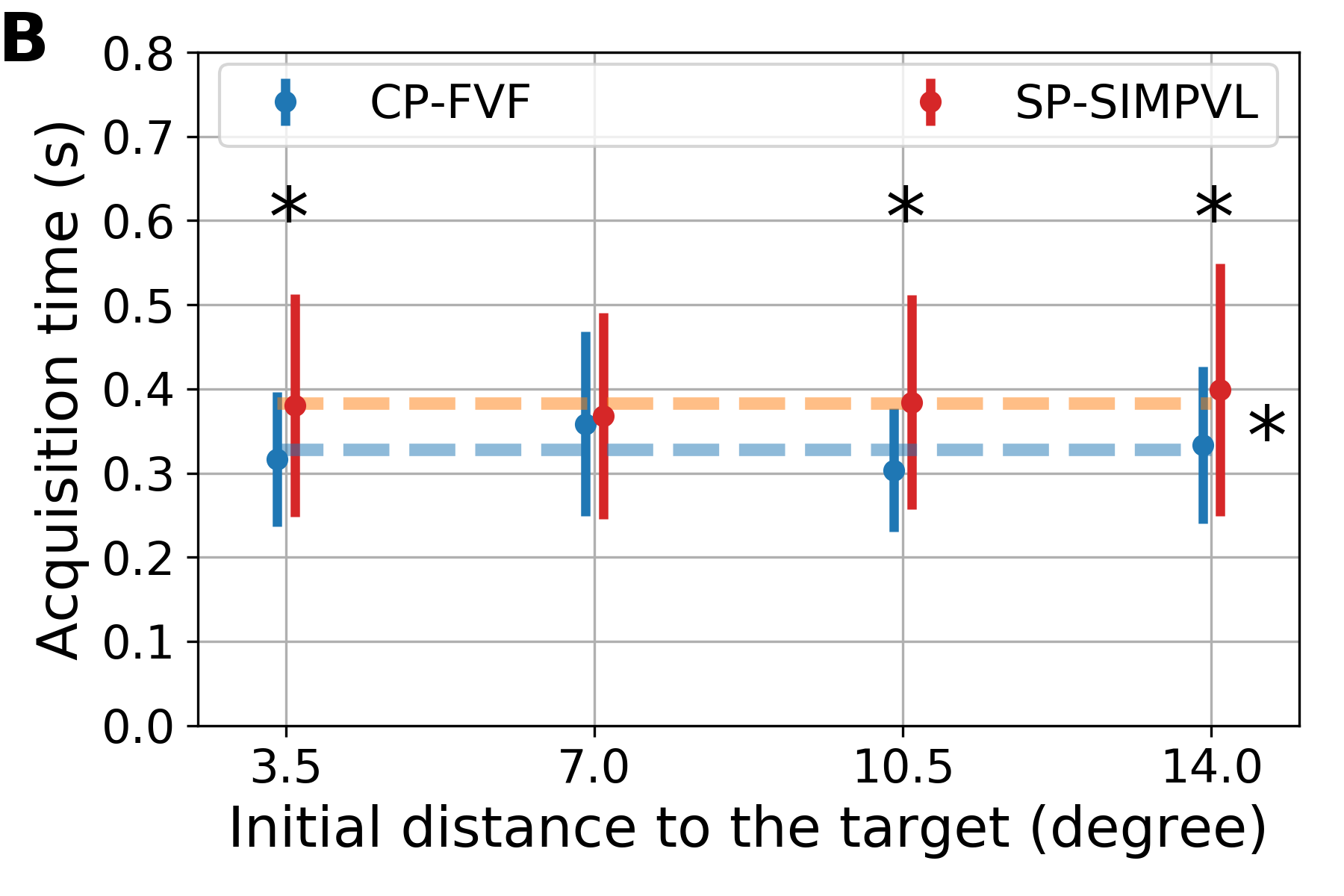} }}%
\qquad
\subfloat{{\includegraphics[width=0.47\linewidth]{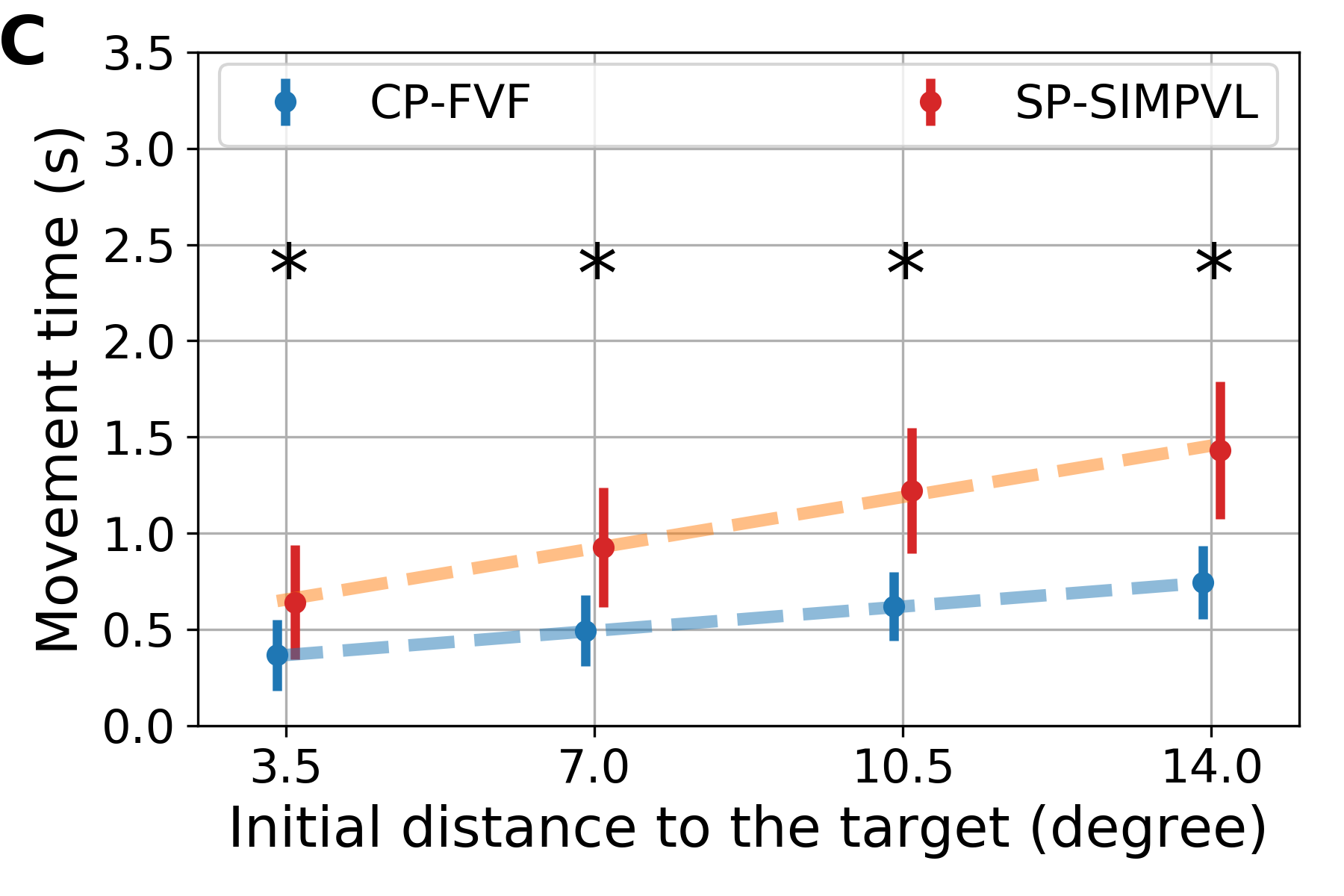} }}%
\qquad
\subfloat{{\includegraphics[width=0.47\linewidth]{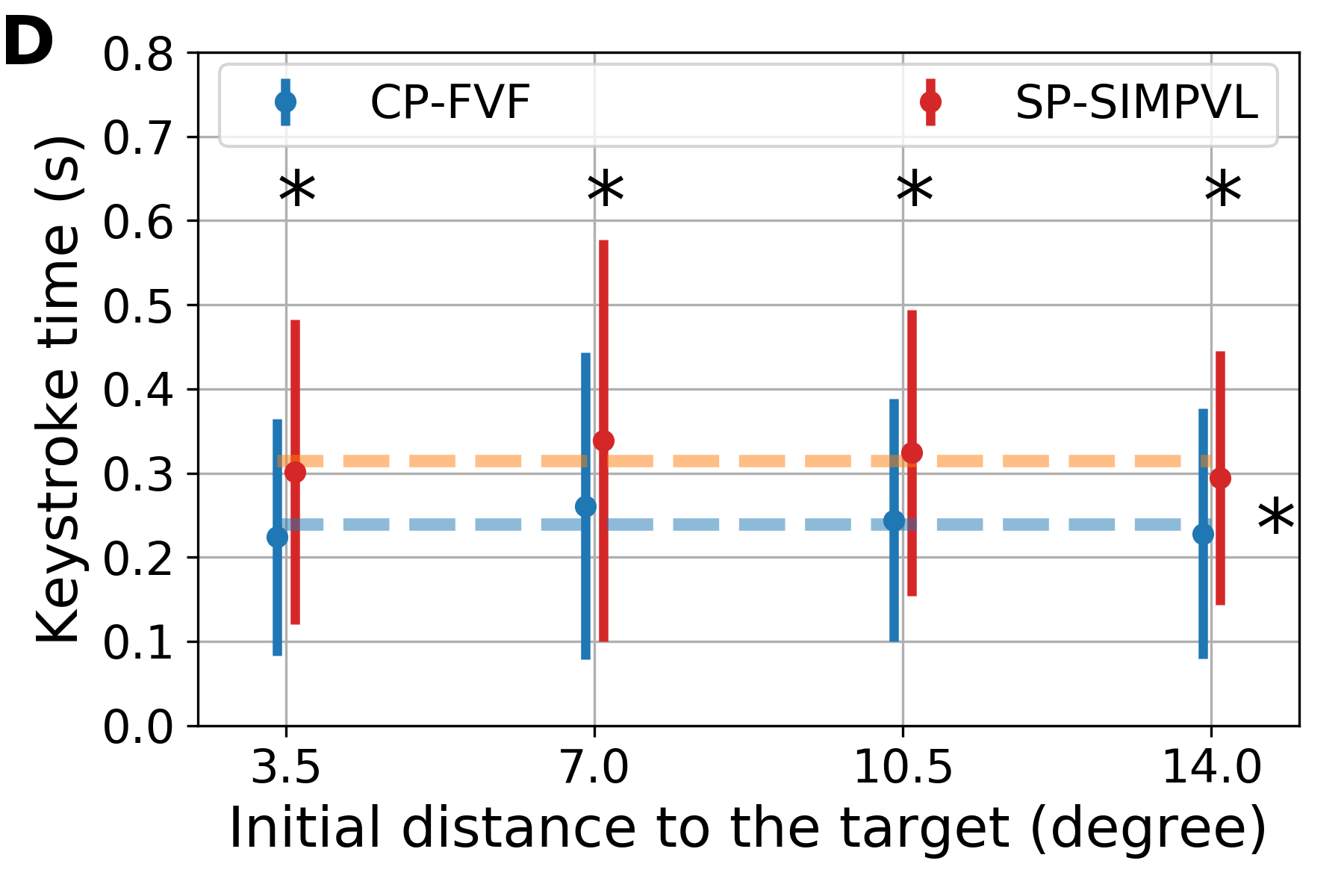} }}%
\caption{
\textbf{A}: the time to complete the task (TCT) against the initial distance between the cursor and the target in the CP-FVF (Crosshair Pointer - Full Visual Field) condition (blue) and the SP-SIMPVL (Sunny Pointer - Simulated Peripheral Vision Loss) condition (red). \textbf{B}: the acquisition time (AT). \textbf{C}: the movement time (MT). \textbf{D}: the time to click on the target (the keystroke time KT). Dashed lines are superimposed to materialize the computed linear regressions of the results. Stars materialize conditions in which the distributions in the CP-FVF and SP-SIMPVL conditions significantly differed according to the Mann-Whitney U test ($p<0.001$).}
\label{fig:results_exp_2_complete}
\end{figure*}

Mean TCTs plotted for the CP-FVF (Crosshair Pointer - Full Visual Field) and the SP-SIMPVL (Sunny Pointer - Simulated Peripheral Vision Loss) conditions against the initial pointer-target distance are shown in figure \ref{fig:results_exp_2_complete}.A. In the SP-SIMPVL condition, the mean TCTs are significantly higher than those observed in the CP-FVF condition by a constant proportion of approximately $50\%$ ($p<0.001$). 

As presented in figure \ref{fig:results_exp_2_complete}.B, the acquisition times do not depend on the pointer-target distance in either condition. The difficulty in estimating the correct direction of movement seems to be independent of the distance of the mouse cursor. In the CP-FVF condition, this suggests that the peripheral vision is good enough to quickly localize the position of the pointer relative to the target, even for a large distance (14$^\circ$). In the SP-SIMPVL condition, this finding suggests that the difficulty in estimating the direction of convergence of the lines does not depend on the proximity of the center of convergence (i.e. the pointer position). 

However, acquisition times are significantly longer in the SP-SIMPVL condition than in the CP-FVF condition by approximately 70ms (390ms VS 320ms). During the acquisition time, the user visually collects information in order to determine the direction and velocity of the mouse cursor in order to move it in optimal fashion toward the target. In the CP-FVF condition, this decision can be made by localizing the mouse cursor in one’s peripheral vision and planning its trajectory relative to the target. On the contrary, in the SP-SIMPVL condition, only the orientation of the lines of the Sunny Pointer and their degree of convergence can be used to estimate the position of the mouse cursor relative to the target. Given the parsimonious visual input available through the restricted aperture, this estimation is not straightforward. The additional 70ms may thus reflect the additional cognitive processing time required by the participant to estimate the direction and the distance of the mouse cursor from the convergence of the "rays" of the pointer. However, even if it is significant, this additional processing time remains remarkably short in view of the difficulty of the task to be performed, again illustrating the effectiveness of the visual system to quickly process complex information \cite{thorpe1996}. 

As shown in figure \ref{fig:results_exp_2_complete}.C, most of the TCT difference results from differences in movement time, with durations significantly longer in the SP-SIMPVL condition by a proportion of approximately $70\%$. This corresponds to additional times of approximately 400, 500, 600 and 700ms for the initial distances of 3.5$^\circ$, 7$^\circ$, 10.5$^\circ$, 14$^\circ$ respectively. 

The keystroke times, presented in figure \ref{fig:results_exp_2_complete}.D, do not depend on the initial distance to the target in either condition. However, a constant and significant additional latency of approximately $90ms$ was measured in the SP-SIMPVL condition. One explanation for this could be that the concentration of the "rays" close to the mouse cursor partly masks the target and thus perturbs the decision as to whether the validation click can be effected or not. This eventuality must be further studied to be confirmed.

\begin{figure*}[ht]\centering
\subfloat{{\includegraphics[width=0.47\linewidth]{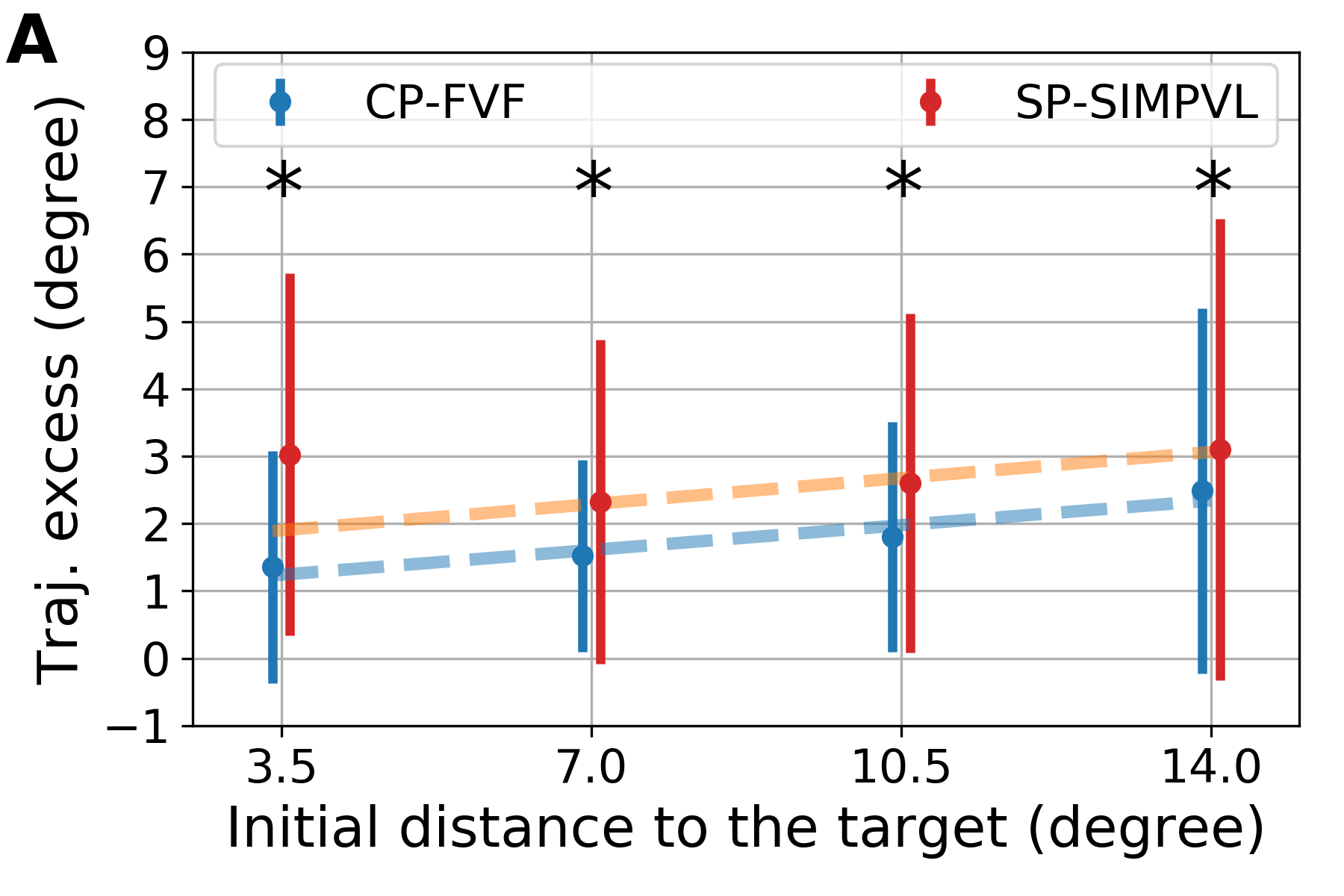} }}%
\qquad
\subfloat{{\includegraphics[width=0.47\linewidth]{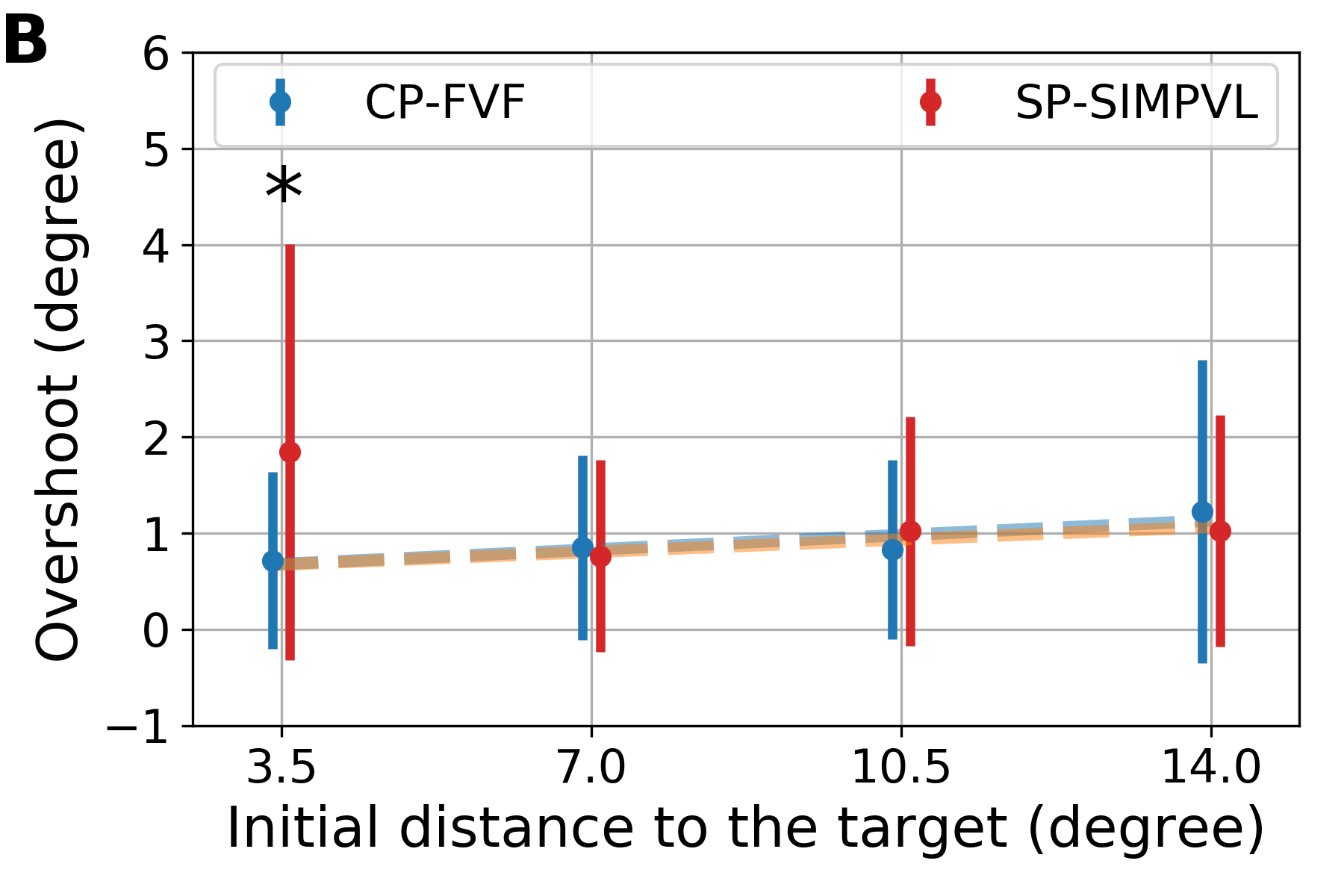} }}%
\qquad
\subfloat{{\includegraphics[width=0.47\linewidth]{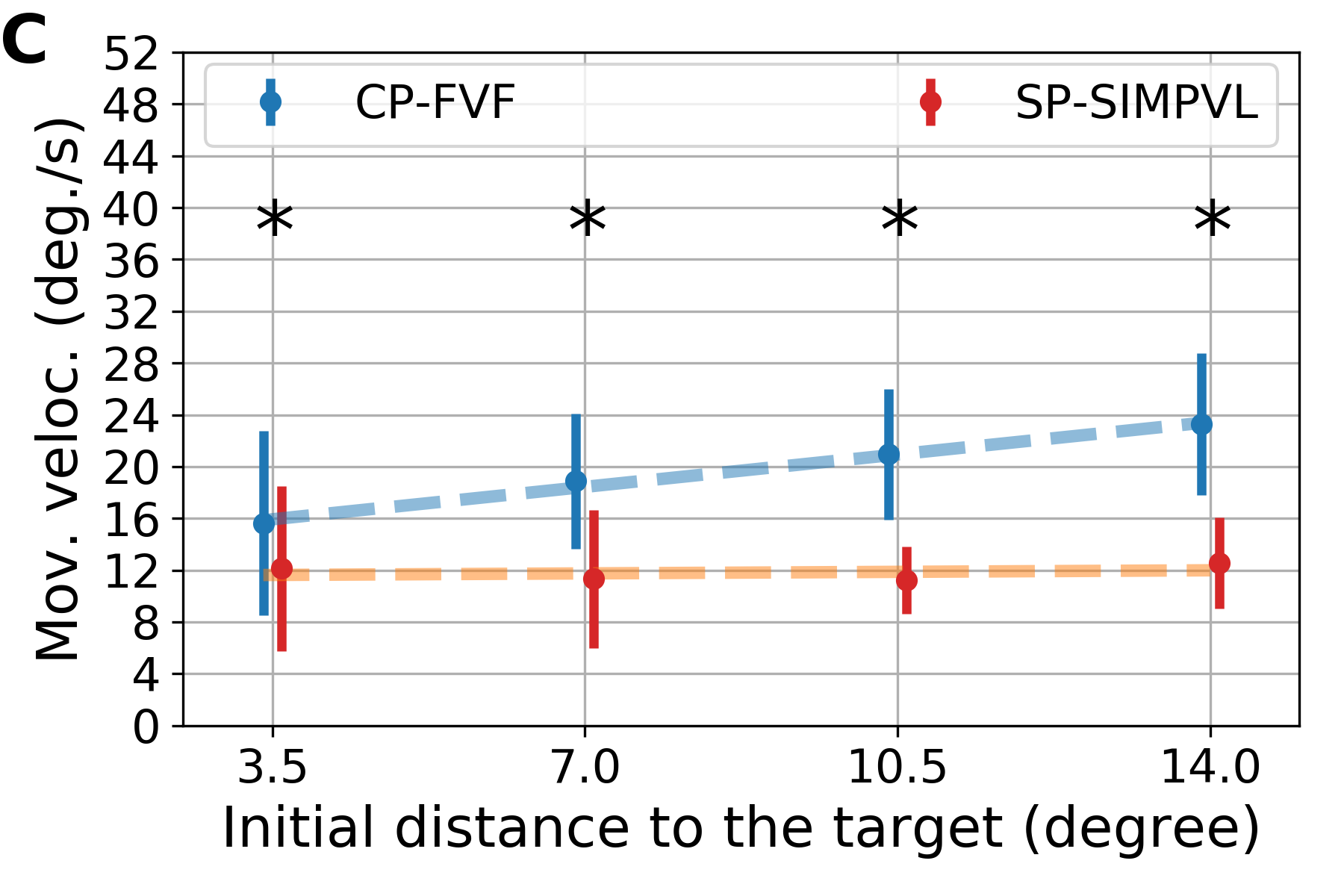} }}%
\qquad
\subfloat{{\includegraphics[width=0.47\linewidth]{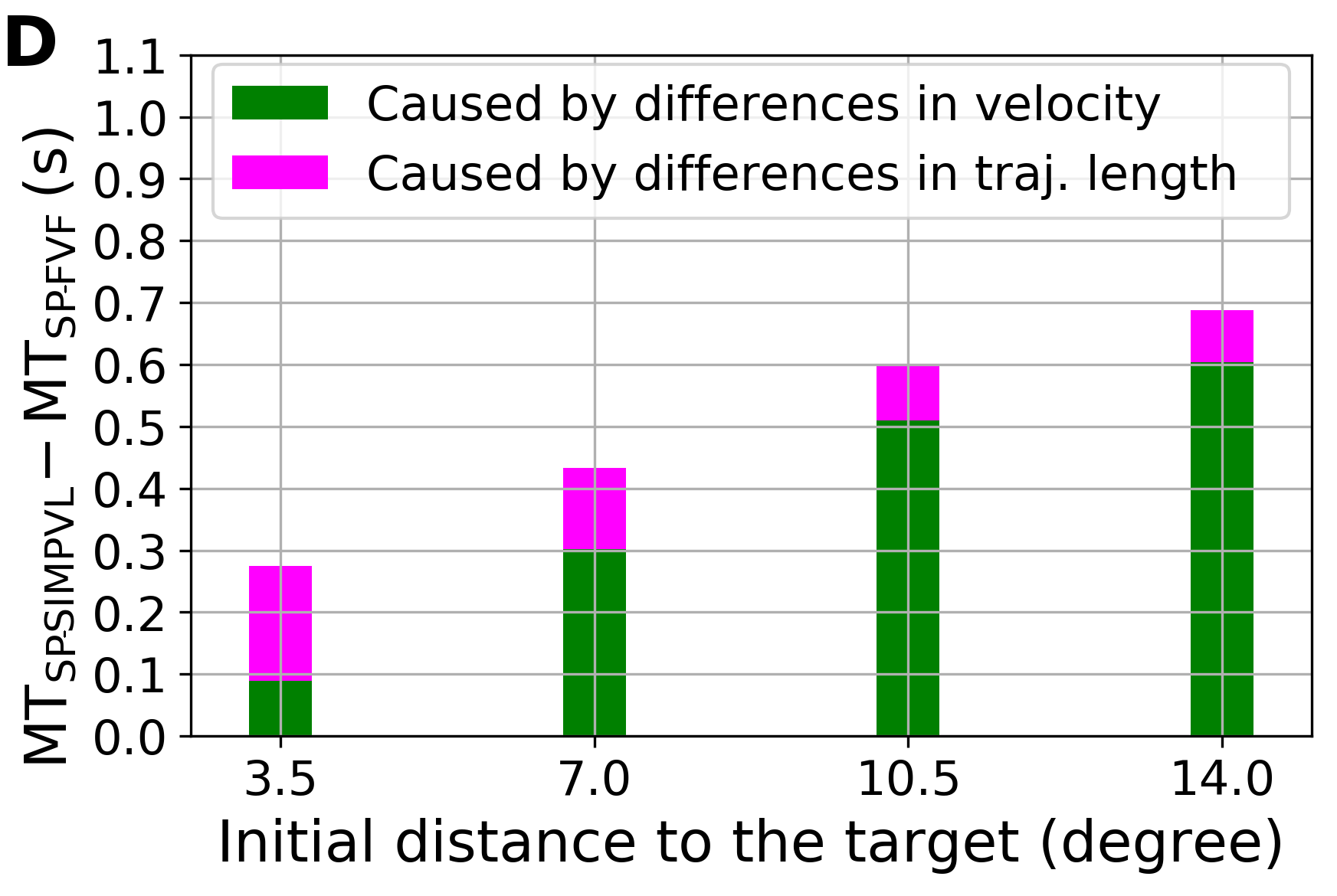} }}%
\caption{\textbf{A}: Trajectory excess computed as the mean length of the trajectories minus the initial distance between the target and the cursor.\textbf{B}: Trajectory excess due to trajectory overshoots. \textbf{C}: Mean velocity of the movements computed as the length of the trajectories divided by the time of movement. \textbf{D}: Differences in movement times between the SP-SIMPVL and the CP-FVF conditions. The differences are partitioned into delays caused by differences in movement velocities and delays caused by differences in trajectory lengths.}
\label{fig:results_diff_exp_2}
\end{figure*}

To better understand the reasons for this longer movement time in the SP-SIMPVL condition, we analyzed the lengths of the paths and the movement velocities of the pointer trajectories. The length of a trajectory was computed by summing the lengths of all the detected moves of the mouse cursor. The trajectory excess is the length of the trajectory path minus the initial pointer-target distance at the beginning of the trial. It thus measures the length of the trajectory that could have been avoided if the control of the mouse cursor direction had been optimal, i.e. if the participant had moved the pointer in a straight trajectory from its initial position to the center of the target. A segment of the cursor trajectory is counted in the overshoot path if its ending point is localized in the half space behind a line passing through the target center point and perpendicular to the segment defined by the target center point and the initial cursor position. 

As shown in \ref{fig:results_diff_exp_2}.A, both in the CP-FVF and the SP-SIMPVL conditions, except for the results in the SP-SIMPVL condition with an initial distance of 3.5$^\circ$, the trajectory excess linearly increases with the initial distance from the target. The paths measured in the SP-SIMPVL condition are longer by a constant amount of approximately 0.7 degrees. Interpreting the orientation of the rays of the Sunny Pointer through the aperture by means of central vision in order to guide the pointer seems thus to be slightly less accurate than using peripheral vision. This illustrates the important role of peripheral vision in trajectory planning, as mentioned in previous works \cite{khan2004}. 

As presented in figure \ref{fig:results_diff_exp_2}.B, the particularly longer path for the initial distance of 3.5$^\circ$ in the SP-SIMPVL condition is mainly caused by more frequent occurrences of overshoots. For the other pointer-target distances, the amount of overshoot in CP-FVF and SP-SIMPVL conditions does not significantly differ and the two profiles show a small linear increase with regards to the pointer-target distance. 

The movement velocity was computed for each trail by dividing the length of the mouse trajectory by the movement time. As shown in figure \ref{fig:results_diff_exp_2}.C, the mean movement velocity in the CP-FVF condition can be accurately approximated by a linear relation with regards to the initial pointer-target distance. We found Velocity=0.71 D + 13 where D is the initial distance from the target. In the SP-SIMPVL condition, we found a near constant velocity equal to $12^\circ/s$. This finding signifies that, with an increasing pointer-target distance, users in the SP-SIMPVL condition do not increase the mean velocity of the move, as would happen in normal pointer use (CP-FVF). The participant might be worried about going too fast and overshooting the target and thus chooses to move the pointer at a velocity that allows him to quickly stop its movement as soon as it enters the visible area around the target. 

Figure \ref{fig:results_diff_exp_2}.D shows the proportional impact of the difference in velocities and the difference in trajectory lengths that cause the differences in MT between the CP-FVF condition and the SP-SIMPVL condition, previously shown in figure \ref{fig:results_exp_2_complete}.C. 
In order to compute these values we first computed the time that the mean trajectory length found in the SP-SIMPVL condition would have taken at the mean velocity observed in the CP-FVF condition. This calculation gives us the delay caused by the differences in lengths. The delay caused by the differences in velocity is thus the movement time difference minus the previously computed delay caused by the differences in lengths. Whereas the longer trajectories observed for short initial distances (3.5$^\circ$) have an important impact (65\%), this impact rapidly decreases for longer distances, 21\%, 14\% and 8\% for 7$^\circ$, 10.5$^\circ$ and 14$^\circ$ respectively. In these last three configurations, a very important proportion of the differences thus stems from the difference in the movement velocity.

\begin{figure*}[ht]\centering
\subfloat{{\includegraphics[width=0.47\linewidth]{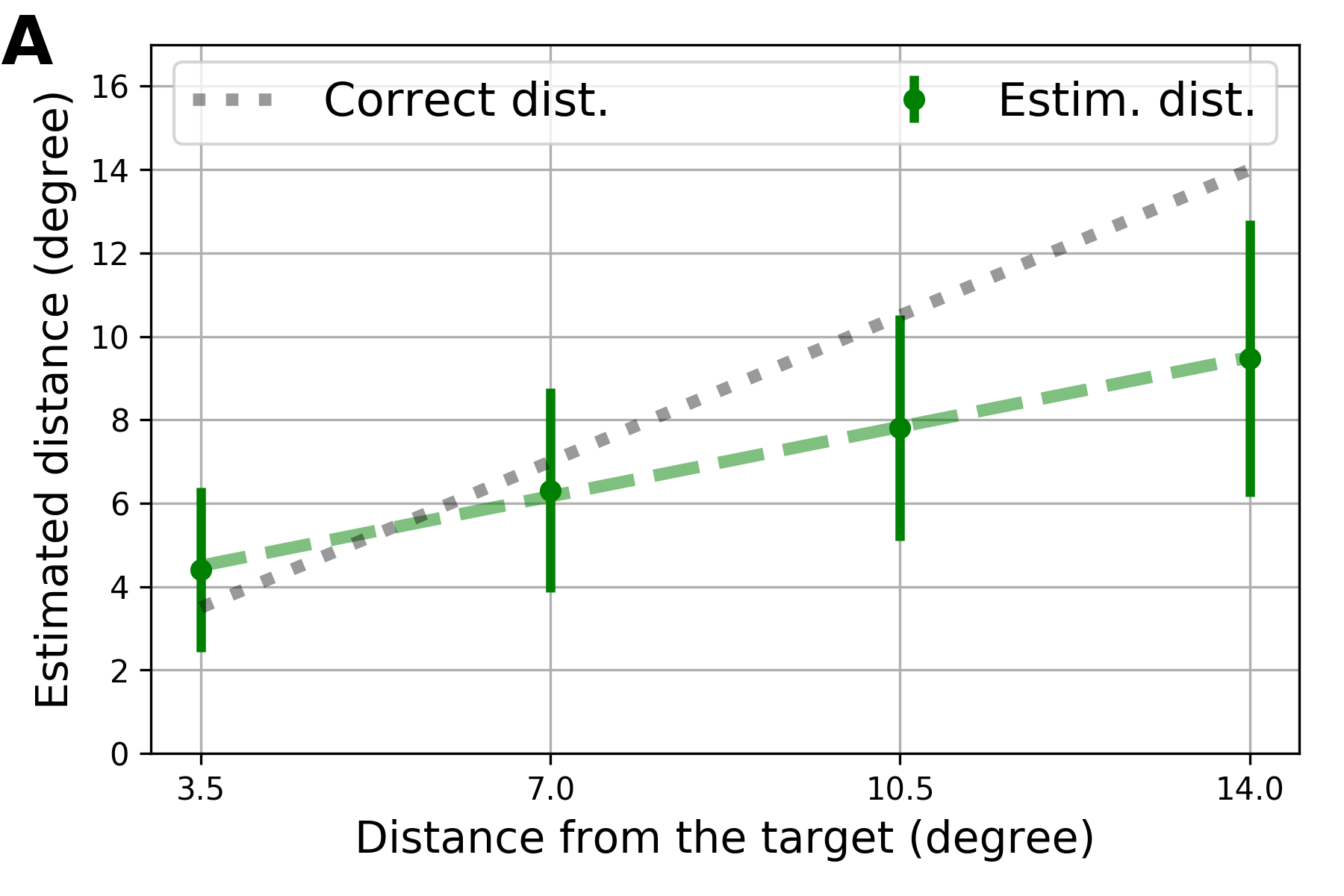} }}%
\qquad
\subfloat{{\includegraphics[width=0.47\linewidth]{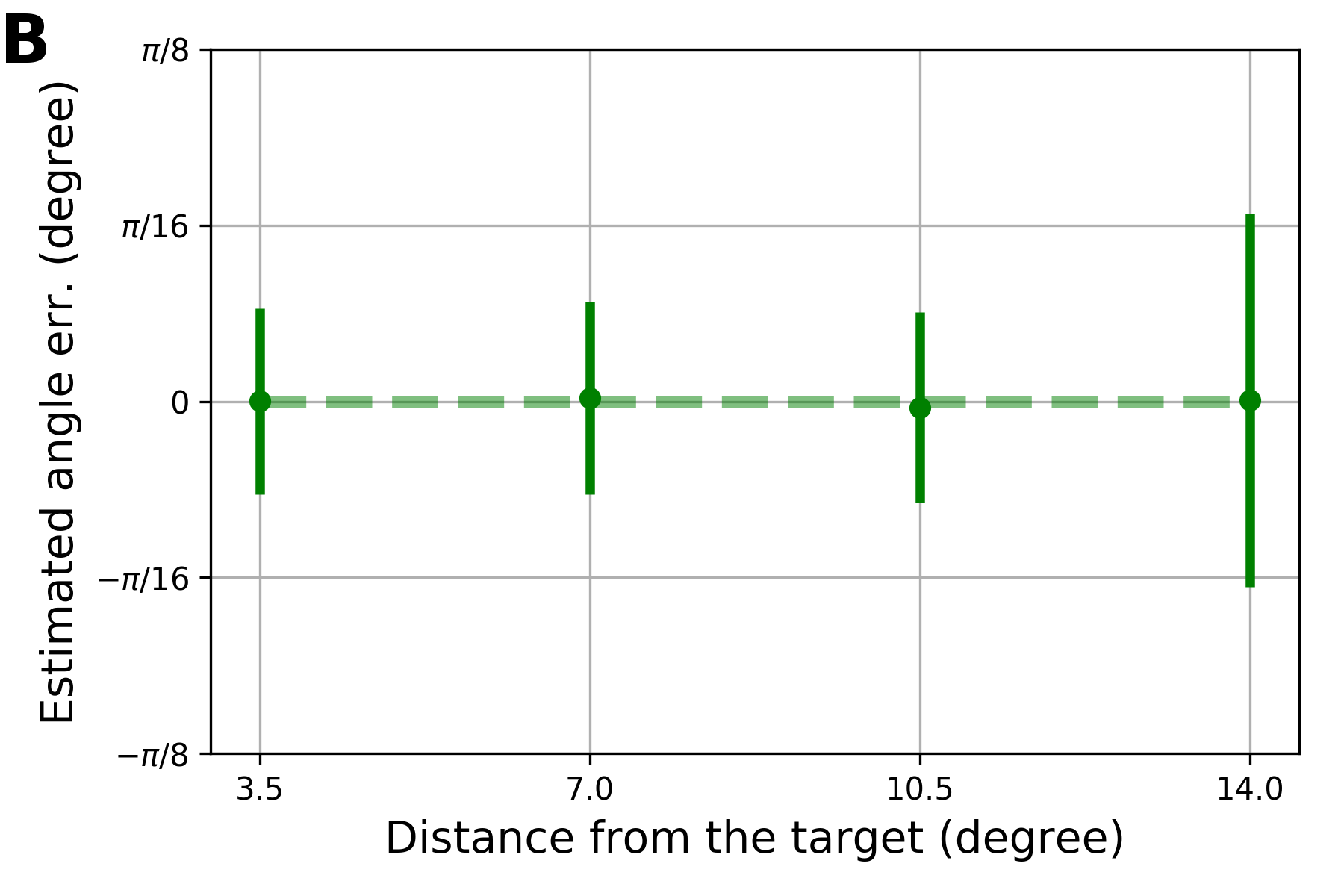} }}%
\caption{\textbf{A}: Estimation of the distance of the convergence point. \textbf{B}: Error in the estimation of the direction of the convergence point.}
\label{fig:results_convergence}
\end{figure*}

Before starting the experiment presented above, we asked each of the 20 participants to complete a preliminary exercise. In this exercise, the mouse cursor was randomly statically placed on the screen for one second with both the Sunny Pointer and the PVL simulation turned on. The user was thus placed in a situation similar to that in the above SP-SIMPVL condition except that the Sunny Pointer was maintained static and therefore no movement toward the target was required. Instead, after one second, both the Sunny Pointer and the PVL simulation vanished and the participant had to click on the screen using a standard mouse cursor at the position where he thought the rays of our pointer displayed on the visible aperture of the screen were converging. For each trial, the convergence point of the Sunny Pointer was randomly placed at various distances and roll angles relative to the center of the screen as in the previously described experiments. 

We compared the distances of the estimated positions to the correct ones. As presented in \ref{fig:results_convergence}.A, the distributions of the estimated distances present an interesting profile. In the case of short distances (3.5$^\circ$), participants overestimated the pointer distance by a mean value of approximately one degree. This overestimation could be the cause of the excess of overshoots seen in figure \ref{fig:results_diff_exp_2}.B for 3.5$^\circ$ in the SP-SIMPVL condition. This overestimation decreases linearly before becoming an underestimation for a convergence point situated at 5.5$^\circ$ from the target. It continues to decrease linearly until finally reaching an underestimation of 4$^\circ$ for a distance of 14$^\circ$ ($\approx29\%$). This miss-estimation can be modeled by means of the following formula: $D^* = -0.47D+2.95$ where $D^*$ is the estimated distance and D is the correct distance. 

The standard deviations for 3.5$^\circ$, 7$^\circ$, 10.5$^\circ$ and 14$^\circ$ are 2$^\circ$, 2.4$^\circ$, 2.6$^\circ$ and 3.1$^\circ$ respectively, which represent 47\%, 39\%, 33\% and 32\% respectively of the mean estimated distances. The standard deviation in the estimated distance can be interpreted as an uncertainty that may explain the flat movement velocity profile shown in figure \ref{fig:results_diff_exp_2}.C in the SP-SIMPVL condition, in which the user moves the mouse slowly due to uncertainty in the estimation of the proximity of the cursor in regards to the target. 

The error in the estimated direction of the convergence is shown in figure \ref{fig:results_convergence}.B. The mean error angle is zero centered, which means that no particular shift is present in the direction estimation. The standard deviation of the error is approximately $\pi/32$ except for 14$^\circ$ with a standard deviation of $\pi/16$. This perhaps reflects an uncertainty in the estimation of the direction that could be the reason for the constant trajectory excess of 0.7 degrees seen in figure \ref{fig:results_diff_exp_2}.A. Nevertheless, the increase in the uncertainty at 14$^\circ$ ($\pi/16$) is not reflected in a particularly larger trajectory excess.

\subsection{Conclusion of the experiment 3}

The use of the Sunny Pointer only partly compensates the absence of peripheral vision in controlling the mouse during a targeting task. This is mainly due to inaccuracy in the estimation of the distance of the pointer based on the convergence of the "rays," resulting in a flat pointer velocity profile. In addition to this velocity effect, the trajectory is a bit longer due to suboptimal orientation of the movements or due to overshoots. And finally, the Sunny Pointer requires more time and probably involves more cognitive load to estimate the correct direction of movement and to click on the target.

\section{Discussion}

Altogether, these results draw a picture that can be interpreted as follows. A person with PVL using a standard mouse takes a long time to retrieve the position of the mouse cursor prior to initiating its movement towards the target on the screen. Moreover, the movement of the pointer towards the target is subject to transitory confusions that may force the user to double check the relative positions of the visual components on the screen during the move, further decreasing efficiency. 
On the contrary, thanks to the Sunny Pointer, visual localization of the mouse pointer on the screen is no longer necessary. Only the focus on the target and an interpretation of the convergence of the "rays" of the pointer above the target are required. This can be effected in less than 400ms and the user can start to move the pointer towards the target a mere fraction of a second after the display. Moreover, the update of the "rays" according to the position of the pointer allows the user to reliably move the pointer in the right direction without transitional confusions. Together, these two contributions of the Sunny Pointer result in a much faster target selection. 

However, although the Sunny Pointer can decrease by a factor of 7 the time required by a person with PVL when using a mouse, this time remains $50\%$ longer than what a person with functional peripheral vision using a standard mouse pointer can achieve. First, the interpretation of the information contained in the visual convergence of the lines of the pointer in a restricted area seems to require an additional cognitive load that translates into approximately an additional 70ms. Secondly, uncertainty surrounding the estimation of the direction of the point of convergence increases the length of the trajectory that the user follows in order to reach the target. This is especially true for short pointer-target distances, for which the user over-estimates the distance of the pointer, which results in frequent overshoots. And thirdly, most of the additional delay comes from a miss-estimation of the distance of the convergence point. As a result, it seems that users choose to ignore their estimations of the distance and prefer to adopt a constant movement velocity, regardless of the distance from the pointer to the target. The adopted velocity seems thus to be a compromise between additional delay caused by movement that is too slow and that caused by overshoots from moving too quickly.

\begin{figure}[ht]\centering
\subfloat{{
    \includegraphics[width=\linewidth]{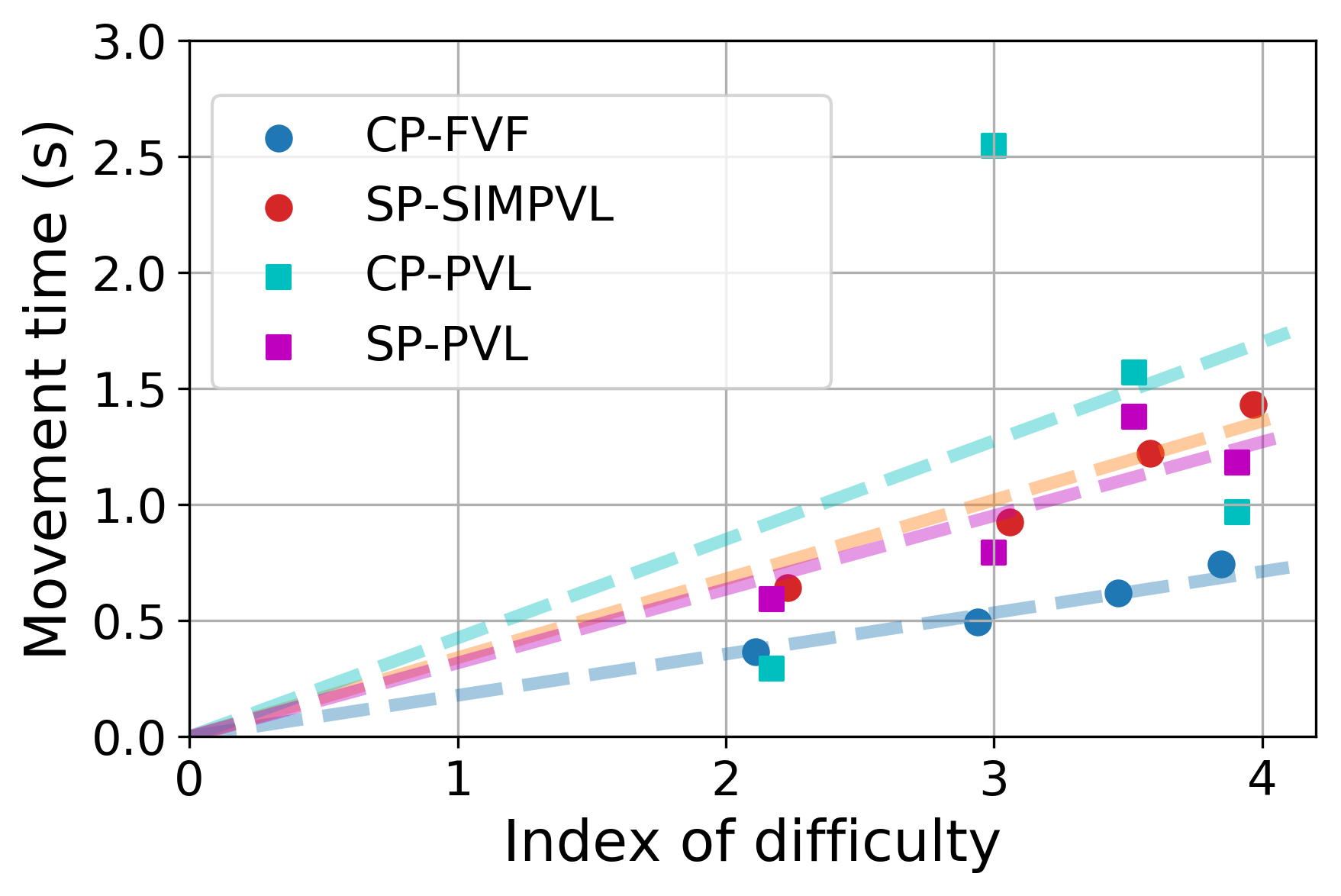}
}}
\vspace{0.1cm}
\subfloat{{
    \begin{tabular}{lrr}
        \toprule
        Condition & Index of perf. & Coef. of deter. ($R^2$)\\
        \midrule
        CP-PVL & $2.35$ & $0.01$ \\
        SP-PVL & $3.15$ & $0.40$ \\
        CP-FVF & $5.62$ & $0.35$ \\
        SP-SIMPVL & $2.94$ & $0.42$ \\
        \bottomrule
    \end{tabular}
}}
\vspace{0.3cm}
\caption{Movement times in the CP-PVL (Crosshair Pointer - Peripheral Vision Loss) and SP-PVL (Sunny Pointer - Peripheral Vision Loss) conditions of experiment 1 and the CP-FVF (Crosshair Pointer - Full Visual Field) and the SP-SIMPVL (Sunny Pointer - Simulated Peripheral Vision Loss) conditions of experiment 2 as a function of the index of difficulty. Linear fits are shown with dashed lines. The summary of the linear fits is presented in the table below.}
\label{fig:results_fitts}
\end{figure}

We plotted in figure \ref{fig:results_fitts} the movement times as a function of the index of difficulty. Despite justified criticism \cite{guiard2011, murata1996}, in order to compared with other studies, we chose Shannon's formulation of the index of difficulty first given in the context of Human-Computer Interaction (HCI) in \cite{mackenzie1992, mackenzie1995} by the formula: $ID=log2(D/W+1)$ where ID is the index of difficulty; D is the distance from the starting point to the center of the target (3.5$^\circ$, 7$^\circ$, 10.5$^\circ$, 14$^\circ$); and W is the width of the target measured along the axis of motion (1$^\circ$). The results were linearly fitted by means of Fitts' law \cite{fitts1954} in which movement time linearly depends on the ID using the equation MT = bID+a where b is a parameter determined by regression analysis. Parameter a (the y-intercept), often considered to be the time to click on the target in paradigms similar to the one we have used \cite{mackenzie1992}, was fixed at $0$ since the keystroke time was excluded from the movement time as explained in \ref{analysis_descr}. In this plot, we thus compare the data presented above in figure \ref{fig:results_exp_2_complete}.C and the data presented in figure \ref{fig:results_exp1_complete}.C on a new x-scale. For purposes of clarity, only the mean values and their associated linear fits are plotted. 

The descriptive data of the fits are summarized in the table under figure \ref{fig:results_fitts}. As previously mentioned, due to confusions during the move of the pointer, the data in the CP-PVL condition shows major variations that disqualify the linear fit as a good predictive model. In the other conditions, Fitts' law can be used as an approximation for the results with a coefficient of determination of approximately 0.4. As expected, the best index of performance is obtained by people with an intact peripheral visual field using a normal mouse pointer (CP-FVF). Next in terms of performance is Jean and his $3.5\%$ of VF (SP-PVL). Arriving in third place is the simulated PVL with the $1.5\%$ of VF (SP-SIMPVL). Thus, the index of performance seems to depend on the size of the VF. A narrower VF might cause greater inaccuracy in the estimation of the position of the pointer relative to the target and thus lead to a slower and more cautious pointer move. This inaccuracy is composed, first, of a shift between the true distance and the mean estimated distance and, second, of an uncertainty in the estimated distance. This possible relationship between uncertainty in the distance estimation and the index of performance requires more experimentation to be confirmed. 

Fitts' law has been used as a predictive model for targeting tasks in the context of HCI for more than 40 years \cite{card1978}. The values we found are in the range of those previously reported: from 2.55 bit/s \cite{epps1986}, 3.2 \cite{boritz1991}, 4.5 bit/s in \cite{mackenzie1991}, 5.7 in \cite{han1990}, up to a value of 10.42 bit/s \cite{card1978}, which is close to the optimal value of 10.56 bit/s found in natural hand movements \cite{fitts1954}. However, we found that the variation in movement time is accounted for by regression equations ($R^2$) to an extent representing approximately $40\%$ whereas the proportion is $70\%$ in \cite{epps1986}, $83\%$ in \cite{card1978} and approximately $90\%$ in \cite{murata1996}. This discrepancy might be due to inter-participant variations. 

How this tool will be used in realistic pointing scenarios will highly depends on each users visual profiles, on their computer setups, and on the specific task they are trying to do. To fit these diversity, the tool we provide proposes many settings in order to customize the graphical display and its ergonomic. It has two activation modes. With the automatic activation mode, the lines appear as soon as a mouse move is detected and follow the subsequent moves of the mouse cursor until the cursor remains static for a period of 0.05s. At this time, the lines vanish until a new mouse move is detected. With the manual mode, the user has to maintain pressed a keys combination to make the lines visible. The Sunny Pointer can be activated, deactivated and closed by means of specific key shortcuts. The number of "rays" radiating from the pointer, the color, the thickness, the transparency, the starting distance of the lines from the mouse pointer as well as their lenght can be adjusted. 

Other visual cues could also be added to simplify the decoding of distance information. The first aim of further developments will be to facilitate the estimation of the pointer distance based on the visual information provided by the Sunny Pointer. Since the Sunny Pointer displays information related to the position of the pointer on the entire area of the screen, any visible area of the screen can contribute to its localization. It would thus be interesting to test if the pointer we have developed might be of help to people with other types of impairments.

\section*{Acknowledgments}
This study was carried out in part with the support of the UNADEV (Union Nationale des Aveugles et Déficients Visuels), the Universit\'{e} de Bourgogne Franche-Comt\'{e} and the CNRS (Centre national de la recherche scientifique). The author would like to thank Jean-Michel Boucheix for the eye tracker, Olivier White for interesting discussions, Jean and all the participants for their enthousiast, Perrine Ambard for precious advices, Céline Tournier for her disponibility and her trust, the associations FIDEV of Lyon and A.I.R of Paris for the use of premisses.

\phantomsection
\bibliographystyle{unsrt}
\bibliography{ambard_sunny_pointer}

\begin{thebibliography}{10}

\bibitem{Rosenholtz_2016}
Ruth Rosenholtz.
\newblock Capabilities and limitations of peripheral vision.
\newblock {\em Annual Review of Vision Science}, 2(1):437--457, 2016.
\newblock PMID: 28532349.

\bibitem{Resnikoff_2004}
Serge Resnikoff, Donatella Pascolini, Ivo Kocur, Ramachandra Pararajasegaram,
  Gopal~P. Pokharel, and Silvio~P. Mariotti.
\newblock Global data on visual impairment in the year 2002.
\newblock {\em Bulletin of the World Health Organization}, 82:844--851, 2004.

\bibitem{Tham2014}
Yih-Chung Tham, Xiang Li, Tien~Yin Wong, Harry~A. Quigley, Tin Aung, and
  Ching-Yu Cheng.
\newblock Global prevalence of glaucoma and projections of glaucoma burden
  through 2040: a systematic review and meta-analysis.
\newblock {\em Ophthalmology}, 121 11:2081--90, 2014.

\bibitem{Quaranta_2016}
Luciano Quaranta, Ivano Riva, Chiara Gerardi, Francesco Oddone, Irene Floriano,
  and Anastasios G.~P. Konstas.
\newblock Quality of life in glaucoma: A review of the literature.
\newblock {\em Advances in Therapy}, 33(6):959--981, 2016.

\bibitem{Ramulu_2009}
P.~Ramulu.
\newblock Glaucoma and disability: which tasks are affected, and at what stage
  of disease?
\newblock {\em Current opinion in ophthalmology}, 20(2):92, 2009.

\bibitem{evans2009}
Keith Evans, S.~K.~Alex Law, John~G. Walt, Patricia Buchholz, and Jan Hansen.
\newblock The quality of life impact of peripheral versus central vision loss
  with a focus on glaucoma versus age-related macular degeneration.
\newblock In {\em Clinical ophthalmology}, volume~3, page 433–45, 2009.

\bibitem{westheimer1982}
Gerald Westheimer.
\newblock The spatial grain of the perifoveal visual field.
\newblock {\em Vision Research}, 22(1):157 -- 162, 1982.

\bibitem{thorpe2001}
Simon Thorpe, Karl R.~Gegenfurtner, Michele Fabre-Thorpe, and Heinrich
  Bülthoff.
\newblock Detection of animals in natural images using far peripheral vision.
\newblock {\em The European journal of neuroscience}, 14:869--76, 10 2001.

\bibitem{larson2009}
Adam~M. Larson and Lester~C. Loschky.
\newblock The contributions of central versus peripheral vision to scene gist
  recognition.
\newblock {\em Journal of Vision}, 9(10):6, 2009.

\bibitem{geisler2006}
Wilson~S. Geisler, Jeffrey~S. Perry, and Jiri Najemnik.
\newblock Visual search: The role of peripheral information measured using
  gaze-contingent displays.
\newblock {\em Journal of Vision}, 6(9):1, 2006.

\bibitem{hooge1999}
Ignace~Th.C Hooge and Casper~J Erkelens.
\newblock Peripheral vision and oculomotor control during visual search.
\newblock {\em Vision Research}, 39(8):1567 -- 1575, 1999.

\bibitem{coeckelbergh2002}
Tanja~R.M Coeckelbergh, Frans~W. Cornelissen, Wiebo~H. Brouwer, and Aart~C.
  Kooijman.
\newblock The effect of visual field defects on eye movements and practical
  fitness to drive.
\newblock {\em Vision Research}, 42(5):669 -- 677, 2002.

\bibitem{khan2004}
Michael~A. Khan, Gavin~P. Lawrence, Ian~M. Franks, and Eric Buckolz.
\newblock The utilization of visual feedback from peripheral and central vision
  in the control of direction.
\newblock {\em Experimental Brain Research}, 158(2):241--251, Sep 2004.

\bibitem{Jacko_1998}
Julie Jacko and Andrew Sears.
\newblock Designing interfaces for an overlooked user group: Considering the
  visual profiles of partially sighted users.
\newblock In {\em Proceedings of ASSETS 1998}, pages 75--77, 01 1998.

\bibitem{Jacko_2000_HCI}
Julie~A. Jacko, Robert H.~Rosa Jr., Ingrid~U. Scott, Charles~J. Pappas, and
  Max~A. Dixon.
\newblock Visual impairment: The use of visual profiles in evaluations of icon
  use in computer-based tasks.
\newblock {\em International Journal of Human–Computer Interaction},
  12(1):151--164, 2000.

\bibitem{proteau2000}
Luc Proteau, Karine Boivin, Stéphane Linossier, and Khémais Abahnini.
\newblock Exploring the limits of peripheral vision for the control of
  movement.
\newblock {\em Journal of Motor Behavior}, 32(3):277--286, 2000.

\bibitem{Fraser_2000}
Julie Fraser and Carl Gutwin.
\newblock A framework of assistive pointers for low vision users.
\newblock In {\em Proceedings of ASSETS 2000}, pages 9--16. ACM Press, 2000.

\bibitem{liu2018}
C.~Liu and R.~Zhao.
\newblock Find the ‘lost’cursor: A comparative experiment of visually
  enhanced cursor techniques.
\newblock In {\em International Conference on Intelligent Computing}, pages
  85--92, August 2018.

\bibitem{Baudisch2003}
P.~Baudisch, E.~Cutrell, and G.~Robertson.
\newblock High-density cursor: a visualization technique that helps users keep
  track of fast-moving mouse cursors.
\newblock In {\em Interact’03}, pages 236--243, September 2003.

\bibitem{Chiang_2005}
Michael~F. Chiang, Roy~G. Cole, Suhit Gupta, Gail~E. Kaiser, and Justin~B.
  Starren.
\newblock Computer and world wide web accessibility by visually disabled
  patients: Problems and solutions.
\newblock {\em Survey of Ophthalmology}, 50(4):394 -- 405, 2005.

\bibitem{Jacko_2000}
Julie Jacko, Armando Barreto, Gottlieb J.~Marmet, Josey Y.~M.~Chu, Holly
  S.~Bautsch, Ingrid U.~Scott, and Robert Rosa.
\newblock Low vision: the role of visual acuity in the efficiency of cursor
  movement.
\newblock In {\em Proceedings of ASSETS 2000}, pages 1--8, 01 2000.

\bibitem{ZoomText}
Vispero.
\newblock {ZoomText}.
\newblock \url{https://www.zoomtext.com/}.

\bibitem{Kline_1995}
Richard~L. Kline and Ephraim~P. Glinert.
\newblock Improving gui accessibility for people with low vision.
\newblock In {\em Proceedings of the SIGCHI Conference on Human Factors in
  Computing Systems}, CHI '95, pages 114--121, New York, NY, USA, 1995. ACM
  Press/Addison-Wesley Publishing Co.

\bibitem{AutoHotKey}
AutoHotkey~Foundation LLC.
\newblock {AutoHotKey}.
\newblock
  \url{http://aiweb.techfak.uni-bielefeld.de/content/bworld-robot-control-software/}.

\bibitem{Hollinworth2011}
N.~Hollinworth and F.~Hwang.
\newblock Cursor relocation techniques to help older adults find 'lost'
  cursors.
\newblock In {\em Proceedings of the SIGCHI Conference on Human Factors in
  Computing Systems}, pages 863--866, May 2011.

\bibitem{sunny_pointer_url}
Université de Bourgogne-Franche~Comté. Ambard~Maxime, LEAD CNRS UMR~5022.
\newblock Sunny pointer.
\newblock
  \url{https://leadserv.u-bourgogne.fr/~sunnypointer/publish/publish.htm}.

\bibitem{Card_1980}
Stuart~K. Card, Thomas~P. Moran, and Allen Newell.
\newblock The keystroke-level model for user performance time with interactive
  systems.
\newblock {\em Commun. ACM}, 23(7):396--410, July 1980.

\bibitem{Smith_2000}
Barton~A. Smith, Janet Ho, Wendy Ark, and Shumin Zhai.
\newblock Hand eye coordination patterns in target selection.
\newblock In {\em Proceedings of the 2000 Symposium on Eye Tracking Research \&
  Applications}, ETRA '00, pages 117--122, New York, NY, USA, 2000. ACM.

\bibitem{thorpe1996}
S.~Thorpe, D.~Fize, C.~Marlot, et~al.
\newblock Speed of processing in the human visual system.
\newblock {\em Nature}, 381(6582):520--522, 1996.

\bibitem{guiard2011}
Yves Guiard and Halla~B. Olafsdottir.
\newblock On the measurement of movement difficulty in the standard approach to
  fitts' law.
\newblock {\em PLOS ONE}, 6(10):1--15, 10 2011.

\bibitem{murata1996}
Atsuo Murata.
\newblock Empirical evaluation of performance models of pointing accuracy and
  speed with a pc mouse.
\newblock {\em International Journal of Human–Computer Interaction},
  8(4):457--469, 1996.

\bibitem{mackenzie1992}
I.~Scott MacKenzie.
\newblock Fitts' law as a research and design tool in human-computer
  interaction.
\newblock {\em Hum.-Comput. Interact.}, 7(1):91--139, March 1992.

\bibitem{mackenzie1995}
I.~Scott MacKenzie.
\newblock Human-computer interaction.
\newblock chapter Movement Time Prediction in Human-computer Interfaces, pages
  483--492. Morgan Kaufmann Publishers Inc., San Francisco, CA, USA, 1995.

\bibitem{fitts1954}
P.M. Fitts.
\newblock The information capacity of the human motor system in controlling the
  amplitude of movement.
\newblock {\em Journal of experimental psychology}, 47(6):381, 1954.

\bibitem{card1978}
Stuart~K. Card, William~K. English, and Betty~J. Burr.
\newblock Evaluation of mouse, rate-controlled isometric joystick, step keys,
  and text keys for text selection on a crt.
\newblock {\em Ergonomics}, 21(8):601--613, 1978.

\bibitem{epps1986}
B.W. Epps.
\newblock Comparison of six cursor control devices based on fitts' law models.
\newblock In {\em Proceedings of the Human Factors and Ergonomics Society
  Annual Meeting}, volume~30, pages 327--331, 1986.

\bibitem{boritz1991}
J.~Boritz, K.~S. Booth, and W.~B. Cowan.
\newblock Fitt's law studies of directional mouse movement.
\newblock In {\em Proceedings of Graphics Interface '91}, GI '91, pages
  216--223, 1991.

\bibitem{mackenzie1991}
I.~Scott MacKenzie, Abigail Sellen, and William A.~S. Buxton.
\newblock A comparison of input devices in element pointing and dragging tasks.
\newblock {\em CHI ’91: Proceedings of the SIGCHI conference on Human factors
  in computing systems}, pages 161--166, 1991.

\bibitem{han1990}
Sung~H. Han, Gerard~C. Jorna, Richard~H. Miller, and Kay~C. Tan.
\newblock A comparison of four input devices for the macintosh interface.
\newblock {\em Proceedings of the Human Factors Society Annual Meeting},
  34(4):267--271, 1990.

\end{thebibliography}


\end{document}